\documentclass[11pt,onecolumn,draftcls,journal]{IEEEtran}
\usepackage{amsmath,amssymb,amsthm, graphicx,float,cite}
\usepackage{algorithm,algorithmic,multirow,bbm,MnSymbol}
\usepackage{setspace,hyperref,array,subfig,relsize}

\hypersetup{
    colorlinks=false,
    pdfborder={0 0 0},
}

\allowdisplaybreaks[1]

\newtheorem{thm}{Theorem}
\newtheorem{lem}{Lemma}
\newtheorem{defn}{Definition}
\newtheorem{prp}{Proposition}
\newtheorem*{pri}{Principle of Optimality}
\newcolumntype{C}[1]{>{\centering\let\newline\\\arraybackslash\hspace{0pt}}m{#1}}

\DeclareMathOperator{\diag}{diag}
\DeclareMathOperator{\tr}{tr}

\DeclareMathOperator{\Sym}{Sym}
\providecommand{\norm}[1]{\lVert#1\rVert}

\begin{document}

\title{\LARGE \bf Active Classification for POMDPs: a Kalman--like State Estimator}
\author{Daphney-Stavroula Zois,$^\star$ \textit{Student Member, IEEE}, Marco Levorato, \textit{Member, IEEE}, and Urbashi Mitra, \textit{Fellow, IEEE}
\thanks{D.-S. Zois and U. Mitra are with the Ming Hsieh Department of Electrical Engineering, University of Southern California, Los Angeles, CA. M. Levorato is with the Computer Science Department, Donald Bren School of Information and Computer Science, University of California, Irvine, CA. e-mail: $\lbrace$zois, ubli$\rbrace$ @usc.edu, marco.levorato@uci.edu. 3740 McClintock Ave., Room 522, Los Angeles, CA 90089-2565. Phone: 213-300-3891}
\thanks{This research has been funded in part by the following grants and organizations: ONR N00014-09-1-0700, NSF CNS-0832186, NSF CCF-0917343,  AFOSR FA9550-12-1-0215, the National Center on Minority Health and Health Disparities (NCMHD) (supplement to P60 MD002254), Nokia and Qualcomm.}
\thanks{Parts of the material in this paper have been previously appeared at ICASSP 2013 and ISIT 2013.}}

\maketitle

\begin{abstract}
The problem of state tracking with active observation control is considered for a system modeled by a discrete--time, finite--state Markov chain observed through conditionally Gaussian measurement vectors. The measurement model statistics are shaped by the underlying state and an exogenous control input, which influence the observations' quality. Exploiting an innovations approach, an approximate minimum mean-squared error (MMSE) filter is derived to estimate the Markov chain system state. To optimize the control strategy, the associated mean--squared error is used as an optimization criterion in a partially observable Markov decision process formulation. A stochastic dynamic programming algorithm is proposed to solve for the optimal solution. To enhance the quality of system state estimates, approximate MMSE smoothing estimators are also derived. Finally, the performance of the proposed framework is illustrated on the problem of physical activity detection in wireless body sensing networks. The power of the proposed framework lies within its ability to accommodate a broad spectrum of active classification applications including sensor management for object classification and tracking, estimation of sparse signals and radar scheduling.
\end{abstract}

\section{Introduction}

\textit{Active classification} refers to the problem of accurately inferring and/or tracking an unknown (usually time--varying) process in an uncertain environment by adaptively exploiting available heterogeneous resources such as different sensor types or actuation. Actively selecting between heterogeneous modes results in different qualitative views of the same process and can lead to significant improvement in estimation performance. As a result, the estimation and control processes are tightly interconnected and to ``maximize the amount of information" on the unknown process at each step, a resource allocation problem must be addressed: \textit{which sensing mode should be employed at each step to provide the next observation?} The active classification problem arises in different forms in a broad spectrum of applications, \emph{e.g.} sensor management for object classification and tracking \cite{WilliamsTSP07,AtiaTSP11,HeroSJ11}, coding with feedback \cite{NaghshvararXiv13}, spectrum sensing \cite{UnnikrishnanTSP10}, amplitude design for channel estimation \cite{RangarajanJSTSP07}, visual search \cite{NaghshvarISIT13}, estimation of sparse signals \cite{HauptSSP12, WeiJSTSP13,MalloyarXiv13}, radar scheduling \cite{KrishnamurthyTSP12} and imaging \cite{RangangaranICASSP05}, graph classification \cite{LigoICASSP13}, health care \cite{ZoisTSP13}, automatic speech recognition \cite{RiccardiTSAP05}, generalized search \cite{NowakIT11} and text, image and video classification and retrieval.

In this paper, the problem of system state tracking with observation control is considered for a system modeled by a discrete--time, finite--state Markov chain. The `hidden' system state is observed through a conditionally Gaussian measurement vector that depends on the underlying system state and an exogenous control input, which shapes the observations' quality. To accurately track the time--evolving system state, we address the joint problem of determining recursive formulae for a structured minimum mean--squared error (MMSE) state estimator and designing a control strategy. Specifically, following an innovations approach, we derive a non--linear approximate MMSE estimator for the Markov chain system state. To obtain a control strategy, we propose a partially observable Markov decision process (POMDP) \cite{BertsekasDPOC05} formulation, where the filter's mean--squared error (MSE) performance serves as the optimization criterion. We also consider the problem of enhancing system state estimates by exploiting both past and future observations and control inputs. More precisely, we derive non--linear approximate MMSE smoothing estimators (fixed--point, fixed--interval, fixed--lag) to acquire improved state estimates and comment on their differences. Finally, we illustrate the framework's performance using real data from a body sensing application.

The current work extends our prior work \cite{ZoisTSP13}, which assumed discrete observations, performed Maximum Likelihood system state detection and employed a worst--case error probability bound as an optimization metric. In fact, the framework proposed herein is much more general and realistic.

We focus on state estimation (versus state detection) for several reasons. First, contrary to our prior work \cite{ZoisTSP13}, the current framework enables a natural joint consideration of estimation and control that allows us to optimize the belief state (MMSE state estimate) \cite{BertsekasDPOC05}, which corresponds to the conditional probability distribution associated with the chain states. As a result, we acquire better belief state estimates, which in turn give rise to high detection accuracy. Second, preliminary results \cite{ZoisGlobalSIP13} suggest that the related cost--to--go functions are concave functions of the predicted belief state implying the potential use of efficient methods for computation. Finally, it is well--known that finding the optimal solution of a POMDP is a computationally intractable problem and thus not suited for large--scale applications. However, we believe that we can significantly accelerate related computations by considering the underlying structure of the related processes and exploiting sparse approximation methods similar to \cite{MaggioniTR06, LevoratoEurasip12}. Our framework constitutes the basis toward addressing these large--scale problems.

The classical Kalman filter (KF) \cite{ChenTR03} along with the fixed--interval \cite{RauchJAIAA65}, fixed--point \cite{MeditchJIC67} and fixed--lag \cite{MeditchJIC67, MooreA73} smoothers are suitable for estimating discrete--time, linear Gauss-Markov systems. Their extensions, \emph{i.e.} the Extended and Unscented KF \cite{ChenTR03}, the Extended Kalman smoother and the Unscented Kalman smoother \cite{Haykin01}, are suitable for general, nonlinear, (non)--Gaussian systems, and they usually adopt a Gaussian approximation for the state distribution, while their performance depends significantly on either some kind of linearization or the careful selection of samples points. To reduce communication costs in sensor network (SN) applications, the problem of state estimation using quantized observations with/without availability of analog measurements has been addressed \cite{RibeiroTSP06, RibeiroCSM10, NerurkarTR12}. For example, the proposed Sign-of-Innovation KF \cite{RibeiroTSP06} and its extensions (see \cite{RibeiroCSM10, NerurkarTR12} and references therein) are based on quantized versions of the measurement innovation and/or real measurements for both Gaussian linear and non-linear dynamical systems. Two well-known approaches for deriving recursive estimators are: 1) the \emph{innovations method} \cite{SegallIT76SPET,Bremaud81}, and 2) the \emph{reference probability method} \cite{ElliottSV95}. The former one defines innovations sequences and exploits martingale calculus to determine the estimator's gain, while the latter introduces a probability measure change to cast the observations independent and identically distributed so as to simplify calculations\footnote{For a very nice survey on the reference probability method, see \cite{ElliottSV95} and references therein.}. MMSE and risk--sensitive\footnote{In contrast to risk--neutral (MMSE) estimation, risk--sensitive estimation penalizes higher--order moments of the estimation error.} estimators have been derived via these methods for discrete--time, finite--state Markov chains observed via discrete observations \cite{SegallIT76REDTPP, ElliottAMO94, BaccarelliSP96} or observations corrupted by white Gaussian noise \cite{ElliottA94, BaccarelliSP96}, but without exerting control. In \cite{KrishnamurthyTSP93}, fixed and sawtooth lag smoothers were derived for the same model as in \cite{ElliottA94}. On the other hand, the work in \cite{PhamdoIT94} proposed an approximate MMSE estimator starting from a maximum $\grave{a}$ posteriori (MAP) detector, while \cite{DeySCL95} derived risk--sensitive recursive estimators for discrete--time, discrete finite--state Markov chains with continuous--valued observations. In contrast to all these works, our proposed estimators are approximate MMSE estimators that build upon the innovations method \cite{SegallIT76SPET, SegallIT76REDTPP, BaccarelliSP96} for discrete-time, finite-state Markov chains observed via \underline{controlled} (\emph{i.e.} observations are actively selected by a controller) conditionally Gaussian measurements.

At this point, we underscore that our work differs from the system state estimation problem in discrete--time, jump Markov linear systems (JMLS) \cite{AckersonTAC70, TugnaitAutomatica82, CostaDTMJLS05}. In these systems, the goal is to estimate the underlying system state given that the system operates in multiple modes, each of which is linear and the switching between them casts the overall system non--linear. The mode change is usually modeled by a discrete--time, finite--state Markov chain, which can be assumed either known \cite{AckersonTAC70, TugnaitAutomatica82, CostaDTMJLS05} or unknown \cite{CostaDTMJLS05} leading to different estimation techniques (see \cite{TugnaitAutomatica82, CostaDTMJLS05} and references therein). In the latter case, the Markov chain true value is usually determined via minimization of the associated posterior detection error probability. In contrast to the above line of work, our system state is a discrete-time Markov chain that we want to estimate, and to achieve this goal, we actively select our measurements. Most prior work in JMLS consists of `passive' approaches, \emph{i.e.} methods that attempt to do the best possible when no control over the observations is exerted. Nonetheless, the problem of designing control sequences to enable discrimination between the multiple modes subject to state and control constraints has also been studied (see for example \cite{BlackmoreTAC08} and references therein). The key differences between \cite{BlackmoreTAC08} and our work are: 1) the control affects \emph{both} state and measurements in \cite{BlackmoreTAC08} in contrast to our case, thus, complicating the derivation of the optimal policy, 2) the control affects state and measurements in a \emph{linear} way contrary to our formulation, and 3) our focus is MSE minimization versus \cite{BlackmoreTAC08}, where the goal is to minimize an detection error probability upper bound.

In the context of SNs, our generic definition of control accommodates the fusion of multiple samples from heterogeneous sensors and thus, generalizes prior frameworks that assume one observation from a single sensor \cite{KrishnamurthyTSP02, GuptaAutomatica06, KrishnamurthyTSP07, NaghshvararXiv12} or $\theta$ samples from $\theta$ sensors \cite{AtiaTSP11,MasazadeCISS12}. Furthermore, our filter's MSE performance is intertwined with the control policy design since the trace of the conditional filtering error covariance matrix constitutes the cost functional of a POMDP, enabling us to focus on the estimation error explicitly. In contrast, prior work usually focuses on generic costs \cite{WuTAC08, KrishnamurthyIT13}, general convex distance measures \cite{KrishnamurthyTSP02, KrishnamurthyTSP07, AtiaTSP11}, detection error probability \cite{NaghshvararXiv12} or performance bounds \cite{GuptaAutomatica06, MasazadeCISS12, NitinawaratTAC13, ZoisTSP13}. Our POMDP proves to be non--standard due to non--linear dependence on the predicted belief state incurring additional complexity in contrast to prior art \cite{WuTAC08,AtiaTSP11, KrishnamurthyIT13} that deals with linear POMDPs. Still, the optimal policy can be derived via stochastic dynamic programming versus \cite{MasazadeCISS12}, where a suboptimal scheme is needed. Determining suboptimal control policies by exploiting techniques such as \cite{KrishnamurthyTSP02, KrishnamurthyTSP07}\footnote{In this case, the authors assume discrete--time, finite--state Markov chains observed via discrete observations.}, where non--linear POMDPs have previously appeared, is out of the scope of the current work. In contrast to \cite{WuTAC08}, where the authors address existence and stability issues for linear quadratic Gaussian control and generic Markovian models, we provide a unified framework of estimation and control for systems modeled by discrete--time, finite--state Markov chains observed via controlled conditionally Gaussian observations. Our work also differs from \cite{KrishnamurthyIT13} since we are interested in equally tracking all system states, not determining when the Markov chain hits a specific target state and terminate tracking. To do so, we choose from a variety of controls versus \cite{KrishnamurthyIT13}, where only two types of controls are considered.

Related problems in statistics and machine learning are the \textit{optimal design of experiments} (OED)  \cite{FedorovTOE72} (part of which is active sequential multihypothesis testing \cite{NaghshvararXiv12, WeiJSTSP13, NitinawaratTAC13}) and \textit{active learning}\footnote{For a nice introduction to active learning and a survey of the corresponding literature, see \cite{SettlesTechRep09} and references therein.}. In OED, the objective is to design experiments that are optimal with respect to some statistical criterion so as to infer an unknown parameterized system. In active learning, the goal is to construct an accurate classifier by utilizing the minimum number of training samples \cite{GolovinNIPS10, SettlesTechRep09, NowakIT11, NaghshvarAllerton12}. This is usually achieved by intelligent adaptive querying, \emph{i.e.} selecting the input patterns for the training process in a statistically optimal way. In comparison to the above, we allow the hypothesis to change with time as a Markov chain and we do not have access to noisy discrete observations but noisy measurement vectors instead. These two characteristics cast the problem more general but harder than the ones already considered in the literature.

Our contributions are as follows. We propose a framework for estimation and control for controlled sensing applications for a very important class of models: discrete--time, finite--state Markov chains observed via \underline{controlled} conditionally Gaussian measurements.
Specifically, we derive recursive formulae for the state estimator, which proves to be \textit{formally} similar to the classical KF, as well as for the three fundamental types of smoothers (fixed--point, fixed--interval, fixed-lag). In addition, we derive a dynamic programming algorithm to derive the optimal control policy, which optimizes the filter's MSE. Last but not least, we provide numerical results validating the performance of the proposed framework on real data from a body sensing application \cite{ZoisTSP13}.

The remainder of the paper is organized as follows. In Section \ref{sec:PS}, we introduce the active state tracking problem by providing the associated stochastic system model and its innovations representation. Next, in Section \ref{sec:MMSESSE}, we derive the Kalman--like estimator and give its MSE performance. We also comment on the differences between the proposed estimator and the standard KF. In Section \ref{sec:OCPD}, we derive the optimal control policy that drives the estimator, while in Section \ref{sec:MMSESSSE}, we derive smoothed estimators in an attempt to acquire more refined system estimates. In Section \ref{sec:EX}, we consider the body sensing application example to illustrate the performance of our framework and we conclude the paper in Section \ref{sec:CON}.

\section{Problem Statement}\label{sec:PS}

We consider a stochastic dynamical system with system state modeled by a discrete-time, finite state Markov chain that evolves in time. The system state is hidden \emph{i.e.} it is observed through a measurement vector that depends both on the underlying state, as well as an exogenous control input selected by a controller. Our goal is to accurately infer the underlying time-evolving system state by shaping the quality of the observations. To this end, we consider the joint problem of determining formulae for the minimum mean-squared error (MMSE) system state estimate from the past observations and controls (\emph{MMSE filter equations}) and the optimal control strategy that drives this estimator. We also consider the problem of acquiring more refined state estimates by exploiting future observations and controls (\emph{MMSE smoother equations}). We begin by introducing the stochastic model of our system.

\subsection{System Model}

We consider a dynamical system, where time is divided into discrete slots and $k = 0, 1, \dotsc$ denotes discrete time. The system state corresponds to a finite-state, first-order Markov chain with $n$ states, \emph{i.e.} $\mathcal{X} = \lbrace \mathbf{e}_{1}, \dotsc  , \mathbf{e}_{n} \rbrace$ with $\mathbf{e}_{i}$ denoting the unit vector with $1$ in the $i$-\emph{th} position and zero everywhere else. The Markov chain is defined on a given probability space $(\Omega, \mathcal{A}, P)$ and is characterized by the transition probability matrix $\mathbf{P}$ with components $P_{j|i} = P(\mathbf{x}_{k+1} = \mathbf{e}_{j} | \mathbf{x}_{k} = \mathbf{e}_{i})$ for $\mathbf{e}_{i}, \mathbf{e}_{j} \in \mathcal{X}$. We assume that these transition probabilities do not change with time, hence the Markov chain is stationary.

The system state $\mathbf{x}_{k}$ is hidden and at each time step, an associated measurement vector $\mathbf{y}_{k}$ is generated. Each such vector follows a multivariate conditionally Gaussian model of the form
\begin{equation}\label{eq:observation_model}
\mathbf{y}_{k} ~ \big |~ \mathbf{e}_{i},\mathbf{u}_{k-1} \sim f(\mathbf{y}_{k}|\mathbf{e}_{i},\mathbf{u}_{k-1}) = \mathcal{N}(\mathbf{m}_{i}^{\mathbf{u}_{k-1}}, \mathbf{Q}_{i}^{\mathbf{u}_{k-1}} ), \forall \mathbf{e}_{i} \in \mathcal{X}
\end{equation}

\noindent
with statistics depending on the underlying system state $\mathbf{x}_{k}$ and a control input $\mathbf{u}_{k-1}$ selected by a controller at the end of time slot $k-1$. We denote the mean vector and covariance matrix of the measurement vector for system state $\mathbf{e}_{i}$ and control input $\mathbf{u}_{k-1}$ as $\mathbf{m}_{i}^{ \mathbf{u}_{k-1} }$ and $\mathbf{Q}_{i}^{ \mathbf{u}_{k-1} }$, respectively. The control input $\mathbf{u}_{k-1}$ can be defined to affect the size of the measurement vector $\mathbf{y}_{k}$ (cf. adaptive estimation of sparse signals in \cite{WeiJSTSP13}), its form, or both and is selected by the controller based on the available information, \emph{i.e.} history of previous control inputs and measurement vectors. We assume that there are a finite number of controls supported by the system, \emph{i.e.} $\mathbf{u}_{k} \in \mathcal{U} = \lbrace \mathbf{u}^{1}, \mathbf{u}^{2}, \dotsc, \mathbf{u}^{\alpha} \rbrace$, and for the moment, we do not consider the case of missing observations.

The above description indicates that we have a \emph{discrete-time dynamical system with imperfect or partially observed state information} \cite{BertsekasDPOC05}, also known as Partially Observable Markov Decision Process (POMDP). Next, we introduce the innovations representation of our system model, which will play a crucial role in the derivation of the filtering and smoothing equations.

\subsection{Innovations Representation of System Model}

We introduce the source sequence of true states $X^{k} = \lbrace \mathbf{x}_{0}, \mathbf{x}_{1}, \dotsc , \mathbf{x}_{k}\rbrace$, the control sequence $U^{k} = \lbrace \mathbf{u}_{0}, \mathbf{u}_{1}, \dotsc , \mathbf{u}_{k}\rbrace$ and the observations sequence $Y^{k} = \lbrace \mathbf{y}_{0}, \mathbf{y}_{1}, \dotsc , \mathbf{y}_{k}\rbrace$. We also define the \textit{global history} $\mathcal{B}_{k} = \sigma \lbrace X^{k}, Y^{k}, U^{k} \rbrace$, the \textit{histories} $\mathcal{B}_{k}^{+} = \sigma \lbrace X^{k+1}, Y^{k}, U^{k} \rbrace$ and $\mathcal{B}_{k}^{-} = \sigma \lbrace X^{k}, Y^{k-1}, U^{k-1} \rbrace$, and the \textit{observation-control history} $\mathcal{F}_{k} = \sigma \lbrace Y^{k}, U^{k-1} \rbrace$, where $\sigma \lbrace z \rbrace$ denotes the $\sigma$-algebra generated by $z$, \emph{i.e.}, the set of all functionals of $z$. Each control input $\mathbf{u}_{k}$ is determined based on the observation-control history $\mathcal{F}_{k}$ \emph{i.e.} $\mathbf{u}_{k} = \eta_{k}(\mathcal{F}_{k})$.

The \textit{innovations sequence} $\lbrace \mathbf{w}_{k} \rbrace$ related to $\lbrace\mathbf{x}_{k}\rbrace$ \cite{SegallIT76SPET} with respect to $\mathcal{B}_{k}$ is defined as
\begin{equation}\label{eq:state_innovations}
\mathbf{w}_{k+1} \doteq \mathbf{x}_{k+1} - \mathbb{E} \lbrace \mathbf{x}_{k+1} | \mathcal{B}_{k} \rbrace,
\end{equation}

\noindent
so that due to the Markov property
\begin{equation}
\mathbb{E} \lbrace \mathbf{x}_{k+1} | \mathcal{B}_{k} \rbrace = \mathbb{E} \lbrace \mathbf{x}_{k+1} | X^{k}, Y^{k}, U^{k} \rbrace = \mathbb{E} \lbrace \mathbf{x}_{k+1} |\mathbf{x}_{k} \rbrace = \mathbf{P}\mathbf{x}_{k}.
\end{equation}

\noindent
Note that the sequence $\lbrace \mathbf{w}_{k} \rbrace$ is a $\lbrace \mathcal{B} \rbrace$--Martingale Difference (MD) sequence \emph{i.e.} it satisfies the following two properties
\begin{equation}
\mathbb{E} \lbrace \mathbf{w}_{k+1} | \mathcal{B}_{k} \rbrace = 0, ~ \forall k \geqslant 0 ~~\text{and}~~ \mathbf{w}_{k+1} \in \mathcal{B}_{k}^{+}, ~\forall k \geqslant 0,
\end{equation}

\noindent
where the last condition implies that $\mathbf{w}_{k+1}$ is a function of $\mathcal{B}_{k}^{+}$. Similarly, the \textit{innovations sequence} $\lbrace \mathbf{v}_{k}\rbrace$ related to the process $\lbrace \mathbf{y}_{k} \rbrace$ \cite{SegallIT76SPET} with respect to $\mathcal{B}_{k}^{-}$ is defined as
\begin{equation}\label{eq:obs_innovations}
\mathbf{v}_{k} \doteq \mathbf{y}_{k} - \mathbb{E} \lbrace \mathbf{y}_{k} | \mathcal{B}_{k}^{-} \rbrace, 
\end{equation}

\noindent
and the following relationship holds
\begin{equation}\label{eq:conditional_mean_00}
\mathbb{E} \lbrace \mathbf{y}_{k} | \mathcal{B}_{k}^{-} \rbrace = \mathbb{E} \lbrace \mathbf{y}_{k} | X^{k}, Y^{k-1}, U^{k-1} \rbrace = \mathbb{E} \lbrace \mathbf{y}_{k} | \mathbf{x}_{k}, \mathbf{u}_{k-1} \rbrace = \mathcal{M}(\mathbf{u}_{k-1}) \mathbf{x}_{k},
\end{equation}

\noindent
where $\mathcal{M}(\mathbf{u}_{k-1}) = [\mathbf{m}_{1}^{\mathbf{u}_{k-1}}, \dotsc, \mathbf{m}_{n}^{\mathbf{u}_{k-1}}]$ and we have exploited the signal model in (\ref{eq:observation_model}). Again, the sequence $\lbrace \mathbf{v}_{k} \rbrace$ is a $\lbrace\mathcal{B^{-}}\rbrace$--MD sequence, \emph{i.e.}
\begin{equation}
\mathbb{E} \lbrace \mathbf{v}_{k} | \mathcal{B}_{k}^{-} \rbrace = 0, ~ \forall k \geqslant 0 ~~\text{and}~~ \mathbf{v}_{k} \in \mathcal{B}_{k}, ~\forall k \geqslant 0.
\end{equation}

Therefore, the \textit{Doob--Meyer decompositions} of $\lbrace \mathbf{x}_{k} \rbrace$ and $\lbrace \mathbf{y}_{k} \rbrace$ with respect to $\mathcal{B}_{k}$ and $\mathcal{B}_{k}^{-}$, respectively, are
\begin{IEEEeqnarray}{rCl}
\mathbf{x}_{k+1} & = & \mathbf{P} \mathbf{x}_{k} + \mathbf{w}_{k+1}, ~k \geqslant 0, \label{eq:system_eqn}\\
\mathbf{y}_{k} & = & \mathcal{M}(\mathbf{u}_{k-1}) \mathbf{x}_{k} + \mathbf{v}_{k},~k \geqslant 1. \label{eq:obs_eqn}
\end{IEEEeqnarray}

\section{System State Estimator }\label{sec:MMSESSE}

In this section, we develop a Kalman--like filter for estimating the discrete-time, finite-state Markov chain system state from past observations and controls based on the theory introduced in \cite{SegallIT76SPET, SegallIT76REDTPP}. Specifically, we derive an approximate MMSE estimate for a point process observed via conditionally Gaussian measurement vectors with statistics nonlinearly influenced by the system state and a non-deterministic control input. We also provide formulae for the filter performance and a comparison between our proposed estimator and the standard KF.

\subsection{Kalman--like Estimator}

We begin by defining the \emph{a posteriori} probability of $\mathbf{x}_{k}$ conditioned on the observation-control history, also known as the \emph{belief state} in the POMDP literature \cite{BertsekasDPOC05}, as
\begin{equation}\label{eq:APP_vector}
\mathbf{p}_{k|k} \doteq \big [p_{k|k}^{1}, \dotsc, p_{k|k}^{n}\big ]^{T} \in \mathcal{P},
\end{equation}

\noindent
where $p_{k|k}^{i} = P(\mathbf{x}_{k} = \mathbf{e}_{i} | \mathcal{F}_{k}), \mathbf{e}_{i} \in \mathcal{X}$, and $\mathcal{P} = \lbrace \mathbf{p}_{k|k} \in \mathbb{R}^{n}: \mathbf{1}_{n}^{T} \mathbf{p}_{k|k} = 1, 0 \leqslant p_{k|k}^{i} \leqslant 1, \forall \mathbf{e}_{i} \in \mathcal{X} \rbrace$. The expected value of $\mathbf{x}_{k}$ conditioned on the observation-control history $\mathcal{F}_{k}$ coincides with $\mathbf{p}_{k|k}$ since
\begin{align}
\mathbf{x}_{k|k} =& ~\mathbb{E} \lbrace \mathbf{x}_{k} | \mathcal{F}_{k} \rbrace = \sum_{i=1}^{n} \mathbf{e}_{i} P(\mathbf{x}_{k} = \mathbf{e}_{i} |\mathcal{F}_{k}) \\=& ~\big [P(\mathbf{x}_{k} = \mathbf{e}_{1} |\mathcal{F}_{k}), \dotsc, P(\mathbf{x}_{k} = \mathbf{e}_{n} | \mathcal{F}_{k})\big ] = \mathbf{p}_{k|k}.
\end{align}

\noindent
From now on, the notation $\mathbf{p}_{k|k}$ will be used to denote the state estimator.

To properly address the problem of optimal nonlinear MMSE estimation, we begin by defining two special sequences: the \textit{estimate innovations sequence} $\lbrace \mu_{k} \rbrace$ and the \textit{observation innovations sequence} $\lbrace \lambda_{k} \rbrace$ (also called the \textit{fundamental MD of the observations}\cite{Bremaud81}) as follows
 \begin{align}
 \mu_{k}& \doteq \mathbf{p}_{k|k} - \mathbf{p}_{k|k-1} = \mathbb{E} \lbrace \mathbf{x}_{k} | \mathcal{F}_{k} \rbrace - \mathbb{E} \lbrace \mathbf{x}_{k} | \mathcal{F}_{k-1} \rbrace \label{eq:estimate_innov_eqn}\\
 \lambda_{k}& \doteq \mathbf{y}_{k} - \mathbf{y}_{k|k-1} = \mathbf{y}_{k} - \mathbb{E} \lbrace \mathbf{y}_{k} | \mathcal{F}_{k-1} \rbrace \label{eq:obs_innov_eqn}.
 \end{align}

\noindent
We can easily prove that both sequences are $\lbrace\mathcal{F}\rbrace$--MD sequences. We note that the innovations sequences in (\ref{eq:estimate_innov_eqn}) and (\ref{eq:obs_innov_eqn}) try to capture the additional information contained in the observation and its impact on the estimate $\mathbf{p}_{k|k}$, similarly to the case of the innovation sequence in the standard KF \cite{ChenTR03}. However, contrary to the standard KF case, where the innovations sequence is a white--noise sequence, herein, the innovations sequence are $\lbrace \mathcal{F}\rbrace$--MD sequences\footnote{Roughly speaking, the MD property can be seen as an ``intermediate" property between independence and uncorrelation\cite{SegallIT76SPET}.}. Next, we state the MD representation theorem \cite{SegallIT76SPET}, which constitutes a very powerful tool for developing recursive nonlinear MMSE Kalman--like estimators by exploiting the innovations sequences introduced above.

\begin{thm}  [Segall \cite{SegallIT76SPET, SegallIT76REDTPP}]\label{thm:MDRT}
The estimate innovations sequence $\lbrace \mu_{k} \rbrace$ is an $\lbrace\mathcal{F}\rbrace$--MD sequence and therefore, it may be represented as a transformation of the observation innovations sequence $\lbrace \lambda_{k} \rbrace$ as
\begin{equation}
\mu_{k} = \mathbf{G}_{k} \lambda_{k},
\end{equation}
\noindent
where $\lbrace \mathbf{G}_{k} \rbrace$ is an $\lbrace\mathcal{F}\rbrace$--adapted sequence that can be computed as follows
\begin{equation}\label{eq:gain_eqn}
\mathbf{G}_{k} = \mathbb{E} \lbrace \mu_{k} \lambda_{k}^{T} | \mathcal{F}_{k} \rbrace \big [ \mathbb{E} \lbrace \lambda_{k} \lambda_{k}^{T} | \mathcal{F}_{k} \rbrace \big ]^{-1}.
\end{equation}
\end{thm}

\noindent
Theorem \ref{thm:MDRT} states that the gain sequence $\lbrace \mathbf{G}_{k} \rbrace$ is an $\lbrace\mathcal{F}\rbrace$--adapted sequence. In general, this implies that the optimal nonlinear MMSE estimator of the sequence $\lbrace \mathbf{x}_{k} \rbrace$ \emph{does not admit a recursive structure}\footnote{Recursiveness is a very desirable property that ensures implementability of estimation in real time and significant memory savings.} since the recursivity property can only be ensured by the predictability property \cite{SegallIT76SPET}. To clarify the difference between adaptability and predictability, we state their respective definitions.

\begin{defn} [Segall \cite{SegallIT76SPET}]
A sequence $\lbrace b_{k} \rbrace$ is said to be $\lbrace\mathcal{F}\rbrace$--adapted if $b_{k}$ is measurable with respect to $\mathcal{F}_{k}, \forall k$.
\end{defn}

\begin{defn} [Segall \cite{SegallIT76SPET}]
A sequence $\lbrace b_{k} \rbrace$ is said to be $\lbrace\mathcal{F}\rbrace$--predictable if $b_{k}$ is measurable with respect to $\mathcal{F}_{k-1}, \forall k$.
\end{defn}

\noindent
There exist some special cases (for more details, see \cite{SegallIT76SPET, SegallIT76REDTPP} and references therein), where it has been successfully shown that the resulting estimator is finite-dimensional as a result of the predictability property being true. Specifically, the gain sequence $\lbrace \mathbf{G}_{k} \rbrace$ is $\lbrace \mathcal{F} \rbrace$--predictable for
\begin{enumerate}
\item[i.] all linear cases in discrete-time including the classical Kalman filter,
\item[ii.] the discrete-time nonlinear case for point processes \cite{SegallIT76SPET, SegallIT76REDTPP, BaccarelliSP96}.
\end{enumerate}

\noindent
At this point, we wish to underscore that in the discrete-time nonlinear case for point processes, the predictability of the gain sequence has been proven for the uncontrolled case. For the controlled case, we can follow the same arguments as in \cite{BaccarelliSP96} and exploit the fact that the control input is measurable with respect to the observation-control history to prove the predictability of the gain sequence. In the case of  discrete-time nonlinear signals in white Gaussian noise, the predictability of the gain sequence has been disproven \cite{SegallIT76SPET}. However, for certain classes of such system, the optimal MMSE estimator still admits a finite-dimensional recursive structure. Specifically, for nonlinear systems characterized by a certain type of Volterra series expansion or state-affine equation, it has been shown that the resulting estimator is recursive and of fixed finite dimension \cite{MarcusTAC79}. In this case, however, a more general theorem by Bremaud and Van Schuppen for very general nonlinear discrete-time systems must be employed \cite{BremaudTR76a, BremaudTR76b}.

The system model equations $(\ref{eq:system_eqn})$ and $(\ref{eq:obs_eqn})$ do not fall into any of the categories above, where the predictability of the gain sequence either holds or fails. Alternatively, direct application of the theorem by Bremaud and Van Schuppen is impossible since their representation constitutes a general representation of the filter equation without any explicit specification for the related terms. We have instead numerically established that the sequence $\lbrace \mathbf{G}_{k} \rbrace$ \underline{cannot} be $\lbrace \mathcal{F} \rbrace$--predictable (see Section \ref{sec:EX}). \textit{Thus, for our problem of interest, the optimal nonlinear Kalman--like MMSE estimator of the sequence $\lbrace \mathbf{x}_{k} \rbrace$ is intrinsically non-recursive (i.e. the resulting estimator is infinite-dimensional).} At this point, inspired by \cite{GalliTC02}, and since a recursive solution is desired within the family of Kalman--like estimators in our case, we \textit{impose} recursivity as a design constraint and use the following approximation\footnote{Note that if the gain sequence is predictable, the approximation symbol in (\ref{eq:pre_gain_eqn}) is replaced with an equality symbol.} 
\begin{equation}\label{eq:pre_gain_eqn}
\mathbf{G}_{k} \approx \mathbb{E} \lbrace \mu_{k} \lambda_{k}^{T} | \mathcal{F}_{k-1} \rbrace \big [ \mathbb{E} \lbrace \lambda_{k} \lambda_{k}^{T} | \mathcal{F}_{k-1} \rbrace \big ]^{-1}.
\end{equation}

\noindent
This approximation along with the Doob--Meyer decompositions (\ref{eq:system_eqn})--(\ref{eq:obs_eqn}) and the definitions in (\ref{eq:estimate_innov_eqn})--(\ref{eq:obs_innov_eqn}) allow us to determine a suboptimal Kalman-type nonlinear MMSE filtered estimator for the Markov chain system state. Namely, exploiting this approximation, we will have that
\begin{equation}\label{eq:Kalman_like_eqn}
\mu_{k} = \mathbf{G}_{k} \lambda_{k}, ~ k \geqslant 0,
\end{equation}

\noindent
where $\mathbf{G}_{k}$ is the time-varying gain given by (\ref{eq:pre_gain_eqn}). Note that for the set of recursive estimators with a Kalman--like structure, the proposed estimator is an optimal MMSE estimator. Theorem \ref{thm:MMSE_estimate} states the recursive formulae for the proposed Kalman--like estimator.

\begin{thm} \label{thm:MMSE_estimate}
The Markov chain system estimate at time step $k$ is recursively defined as
\begin{equation}\label{eq:filter_eqn}
\mathbf{p}_{k|k} = \mathbf{p}_{k|k-1} + \mathbf{G}_{k} [\mathbf{y}_{k} - \mathbf{y}_{k|k-1} ],~ k \geqslant 0
\end{equation}

\noindent
with
\begin{align}
\mathbf{p}_{k|k-1} &= \mathbf{P} \mathbf{p}_{k-1|k-1}, \label{eq:predictor}\\
\mathbf{y}_{k|k-1} &= \mathcal{M}(\mathbf{u}_{k-1}) \mathbf{p}_{k|k-1},\\
\mathbf{G}_{k} &= \mathbf{\Sigma}_{k|k-1} \mathcal{M}^{T}(\mathbf{u}_{k-1})(\mathcal{M}(\mathbf{u}_{k-1}) \mathbf{\Sigma}_{k|k-1}\mathcal{M}^{T}(\mathbf{u}_{k-1}) + \widetilde{\mathbf{Q}}_{k})^{-1},
\end{align}

\noindent
where $\mathbf{x}_{0|-1} = \pi$, and $\pi$ is the initial distribution over the system states, $\mathbf{\Sigma}_{k|k-1}$ is the conditional covariance matrix of the prediction error and $\widetilde{\mathbf{Q}}_{k} = \sum_{i=1}^{n} p_{k|k-1}^{i} \mathbf{Q}_{i}^{\mathbf{u}_{k-1}}$.
\end{thm}

\begin{IEEEproof}
For proof, see Appendix \ref{app:pfThm_MMSE_estimate}.
\end{IEEEproof}

At this point, we underscore that even though the proposed filter is \emph{formally} similar to the classical KF, it is not a standard KF. In fact, the gain $\mathbf{G}_{k}$ \textit{depends on the observations} and the resulting filter is non-linear in contrast to the classical KF, which constitutes a linear filter. Furthermore, since no constraint is imposed on the individual components of $\mathbf{p}_{k|k}$, there is no guarantee that they lie on the $[0,1]$ interval. To overcome this issue without incorporating additional constraints that may challenge the determination of a solution to our problem, we adopt the approach of \cite{BaccarelliSP96}, \emph{i.e.} apply a suitable memoryless (linear or nonlinear) transformation of $\mathbf{p}_{k|k}$ to ensure feasible solutions are determined.

\subsection{Filter Performance}

The mean-squared error (MSE) performance of the filter in (\ref{eq:filter_eqn}) is intertwined with the \textit{conditional filtering error covariance matrix}, which can be directly computed as follows
\begin{equation}\label{eq:ccmfe}
\mathbf{\Sigma}_{k|k} \doteq \mathbb{E} \lbrace (\mathbf{x}_{k} - \mathbf{p}_{k|k}) (\mathbf{x}_{k} - \mathbf{p}_{k|k})^{T} | \mathcal{F}_{k} \rbrace = \diag ( \mathbf{p}_{k|k} ) - \mathbf{p}_{k|k} \mathbf{p}_{k|k}^{T}.
\end{equation}

\noindent
Similarly, the MSE performance of the predictor in  (\ref{eq:predictor}) is characterized by the \textit{conditional prediction error covariance matrix}, which can be again computed as
\begin{equation}
\mathbf{\Sigma}_{k|k-1} \doteq \mathbb{E} \lbrace (\mathbf{x}_{k} - \mathbf{p}_{k|k-1}) (\mathbf{x}_{k} - \mathbf{p}_{k|k-1})^{T} | \mathcal{F}_{k-1} \rbrace = \diag ( \mathbf{p}_{k|k-1} ) - \mathbf{p}_{k|k-1} \mathbf{p}_{k|k-1}^{T}.
\end{equation}

\noindent
Both previous equations are directly obtained from their definitions and from the fact that the states of the Markov chain constitute the standard orthonormal basis. We can also derive the same recursive equations for the conditional error covariance matrices presented in \cite{BaccarelliSP96} and repeated below for reasons of completeness. Specifically, the conditional filtering error covariance matrix can also be expressed as
\begin{equation}\label{eq:rec_fe}
\mathbf{\Sigma}_{k|k} = \mathbf{\Sigma}_{k|k-1} + \diag ( \mu_{k} ) - \mathbf{G}_{k} \lambda_{k} \lambda_{k}^{T} \mathbf{G}_{k}^{T} - 2\Sym( \mathbf{p}_{k|k-1}\mathbf{G}_{k}^{T}\lambda_{k}^{T}),
\end{equation}

\noindent
where $\Sym( \mathbf{B} ) = \frac{1}{2}(\mathbf{B} + \mathbf{B}^{T})$ and the conditional prediction error covariance matrix as
\begin{equation}\label{eq:rec_pe}
\mathbf{\Sigma}_{k|k-1} = \mathbf{P} \mathbf{\Sigma}_{k-1|k-1} \mathbf{P}^{T} + \diag ( \mathbf{p}_{k|k-1} ) - \mathbf{P} \diag ( \mathbf{p}_{k-1|k-1} ) \mathbf{P}^{T},
\end{equation}

\noindent
which should be initialized as follows
\begin{equation}
\mathbf{\Sigma}_{1|0} = \text{diag} \lbrace \pi \rbrace - \pi \pi^{T}.
\end{equation}

\noindent
Both recursive equations are \emph{formally} similar to the Riccati equation for the standard KF \cite{BertsekasDPOC05}. Furthermore, the specific form of (\ref{eq:rec_fe}) and (\ref{eq:rec_pe}) reveal the filter gain's dependence on the observations.

\subsection{Standard KF versus Markov chain Kalman--like filter}

\begin{figure}[h!]
\centering
\subfloat[Interconnection of basic state-variable system model and standard KF.]
{\label{fig:KF}
\includegraphics[width=0.7\textwidth]{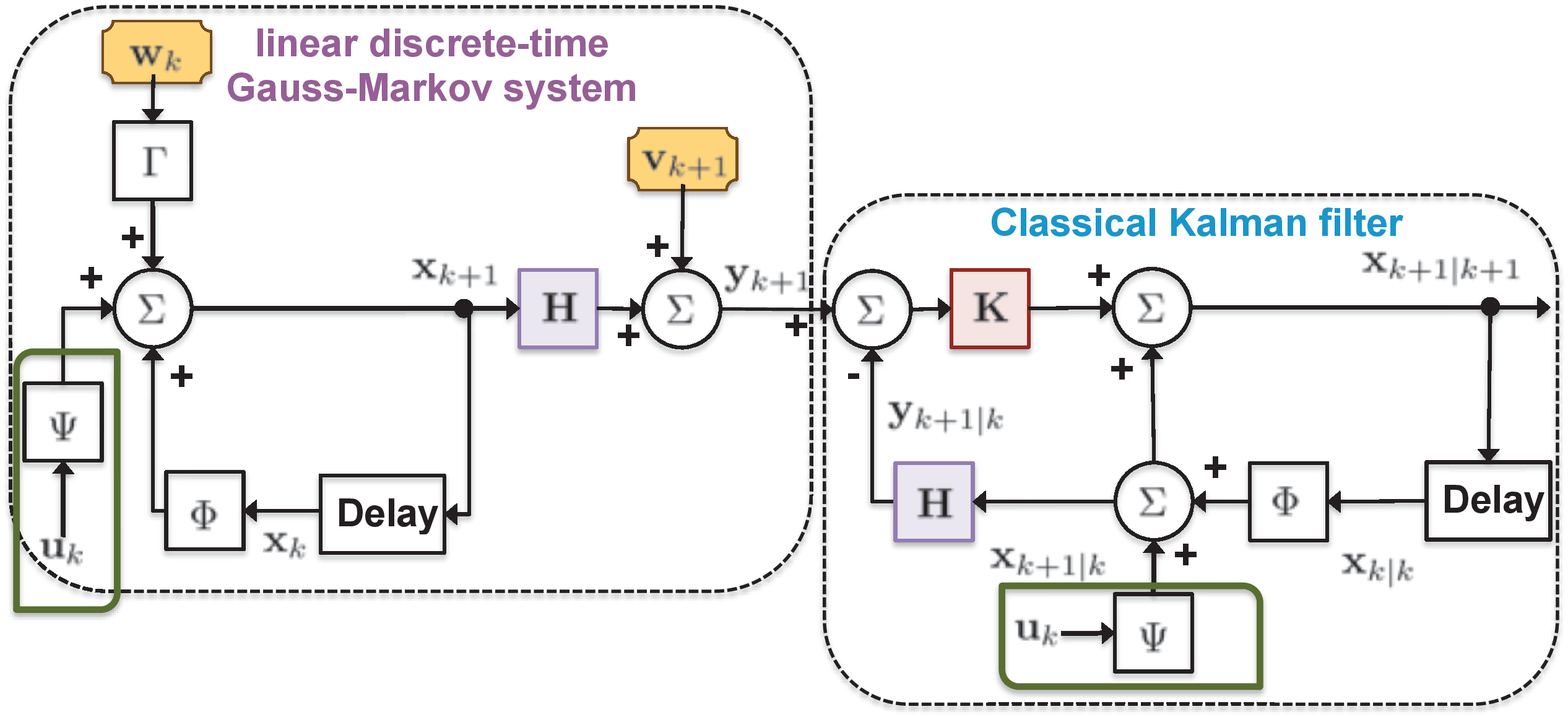}}\\
\subfloat[Interconnection of proposed system model and  Kalman--like filter]
{\label{fig:Markov_chain_KF}
\includegraphics[width=0.7\textwidth]{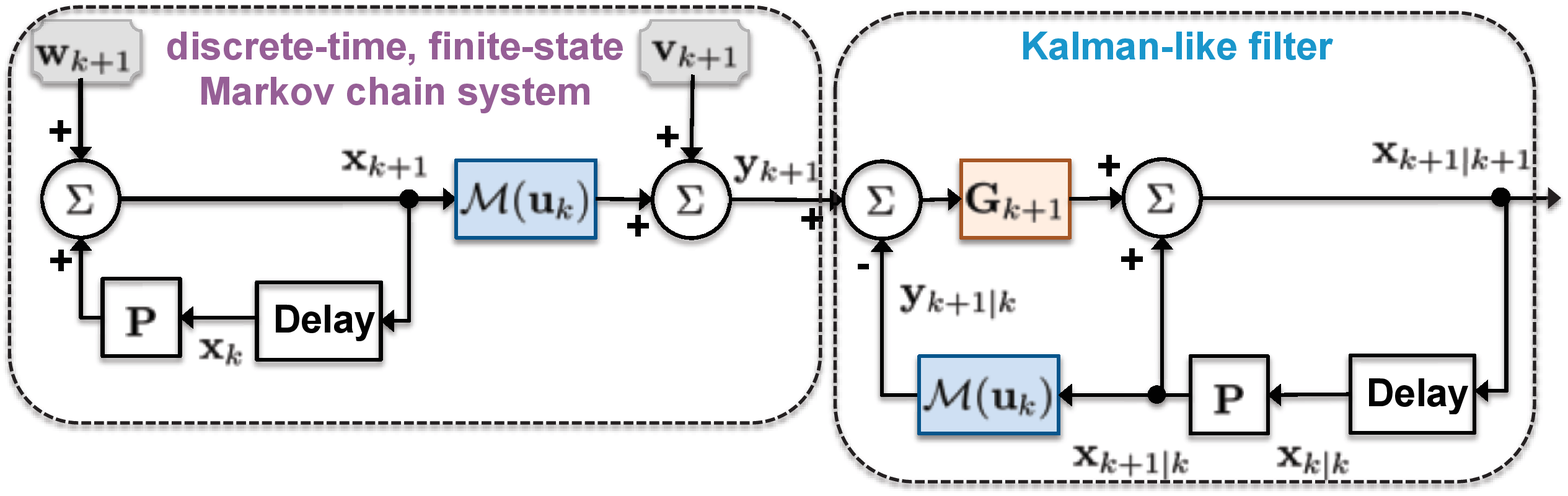}}
\caption{Interconnection of system block diagram and MMSE estimator block diagram.}
\label{fig:models_Kalman_filters}
\end{figure}

In this section,  we comment on the similarities and differences of the mean-squared filtered estimator given in Theorem \ref{thm:MMSE_estimate} and the standard KF \cite{ChenTR03}. Figure~\ref{fig:models_Kalman_filters} shows the system model and the corresponding filter as well as the interconnection between them.

Comparing the block diagrams given in Fig.~\ref{fig:KF} and \ref{fig:Markov_chain_KF} respectively, we observe that their formal structure is similar, \emph{e.g.} both filters contain within their structures a model of the plant, processing is done following the same sequence of steps, \emph{etc}. The main difference between the two estimators lies mainly on the underlying dynamical system they assume. The system model for the standard KF, shown in Fig.~\ref{fig:KF}, assumes that (i) the state and measurement equations are linear, (ii) $\lbrace \mathbf{x}_{k} \rbrace$ is a Gauss-Markov sequence since all related processes have Gaussian distributions, and (iii) the control input linearly influences the system state. In contrast, our system model, shown in Fig.~\ref{fig:Markov_chain_KF}, assumes that (i) the state and measurement equations include non-linear terms, (ii) $\lbrace \mathbf{x}_{k} \rbrace$ is a discrete-time, finite-state Markov chain and the associated measurements conditioned on the system state and the control input are Gaussian, and (iii) the control input influences the measurements in a non-linear fashion. Furthermore, in the standard KF setting, the role of the control is to affect the system state evolution in contrast to our case, where the control affects only the measurements' quality. Another important difference between the two estimators relates to the filter gain in the sense that the KF gain does not dependent on the measurements as it is the case with the gain of our estimator. A direct outcome of this dependence in conjunction with our system model is that our proposed estimator constitutes a non-linear filter opposed to the standard KF, which is a linear filter. Finally, in the standard KF setting, the conditional distribution of the system state proves to be Gaussian with the system state estimate being the conditional mean and the conditional filtering error covariance matrix being the conditional covariance matrix. However, in our setting, this distribution coincides with the system state estimate.

\section{Optimal Control Policy Design}\label{sec:OCPD}


We consider the active state tracking problem introduced in Section \ref{sec:PS}, where the information available to the controller at time $k$ consists of the observation--control history defined earlier as
\begin{equation}\label{eq:obs_history}
\begin{split}
\mathcal{F}_{k}& = \sigma \lbrace \mathbf{y}_{0}, \mathbf{y}_{1}, \dotsc , \mathbf{y}_{k}, \mathbf{u}_{0}, \mathbf{u}_{1}, \dotsc , \mathbf{u}_{k-1} \rbrace,\quad k = 1,2,\dotsc, L, \\
\mathcal{F}_{0}& = \sigma \lbrace \mathbf{y}_{0} \rbrace.
\end{split}
\end{equation}

\noindent
We are interested in determining an admissible control policy $\gamma = \lbrace \eta_{0}, \eta_1, \dotsc, \eta_{L-1} \rbrace$ \cite{BertsekasDPOC05} that minimizes the cost function
\begin{equation}
J_{\gamma} = \underset{\mathbf{y}_{0},\mathbf{y}_{1},\dotsc,\mathbf{y}_{L}}{\mathbb{E}} \bigg \lbrace \sum_{k=1}^{L} \tr{\big (\mathbf{\Sigma}_{k|k}(\mathbf{y}_{k},\mathbf{u}_{k-1}) \big )} \bigg \rbrace
\end{equation}

\noindent
subject to the system equation (\ref{eq:system_eqn}) and the observations equation (\ref{eq:obs_eqn}) with $L$ denoting the horizon length and $\mathbf{u}_{k} = \eta_{k}(\mathcal{F}_{k})$. The term $\tr{\big ( \mathbf{\Sigma}_{k|k}(\mathbf{y}_{k},\mathbf{u}_{k-1}) \big )}$ denotes the trace of the conditional filtering error covariance matrix with its dependence on the measurement vector $\mathbf{y}_{k}$ and the control input $\mathbf{u}_{k-1}$ stated explicitly. Thus, we have the following \emph{finite horizon, partially observable stochastic control problem}
\begin{equation}
\min_{\mathbf{u}_{0},\mathbf{u}_{1},\dotsc,\mathbf{u}_{L-1}} J_{\gamma}.
\end{equation}

\noindent
In contrast to standard problems of this type \cite{BertsekasDPOC05, SpeyerChung08}, we note that our cost function is defined with respect to the observations, not the system states. This fact along with the definition of our cost function influences the form of the solution. To determine the optimal policy, we exploit the ideas in \cite{BertsekasDPOC05}, \emph{i.e.}

\begin{enumerate}
\item[i.] We first reformulate our problem as a perfect state information problem using $\mathcal{F}_{k-1}$ as the new system state, and derive the corresponding dynamic programming (DP) algorithm.
\item[ii.] Next, we determine a sufficient statistic for control purposes and derive a simplified DP algorithm, which solves for the optimal control policy.
\end{enumerate}

\subsection{Perfect State Information Reformulation \& DP Algorithm}

In this section, we reduce our problem from imperfect to perfect state information and then, we derive the corresponding DP algorithm. From the observations--control history definition in (\ref{eq:obs_history}), we observe that
\begin{equation}
\mathcal{F}_{k} = (\mathcal{F}_{k-1},\mathbf{y}_{k},\mathbf{u}_{k-1}), \quad k = 1, 2, \dotsc, L-1,\quad \mathcal{F}_{0} = \sigma \lbrace \mathbf{y}_{0} \rbrace. 
\end{equation}

\noindent
The above equations can be seen as the evolution of a system with system state $\mathcal{F}_{k-1}$, control input $\mathbf{u}_{k-1}$, and random ``disturbance" $\mathbf{y}_{k}$. Furthermore, we have that $p(\mathbf{y}_{k} | \mathcal{F}_{k-1}, \mathbf{u}_{k-1}) = p(\mathbf{y}_{k} | \mathcal{F}_{k-1}, \mathbf{u}_{k-1}, \mathbf{y}_{0}, \mathbf{y}_{1}, \dotsc, \allowbreak \mathbf{y}_{k-1})$, since by definition, $\mathbf{y}_{0}, \mathbf{y}_{1}, \dotsc, \mathbf{y}_{k-1}$ are part of $\mathcal{F}_{k-1}$, and thus, the probability distribution of $\mathbf{y}_{k}$ depends explicitly only on the state $\mathcal{F}_{k-1}$ and control input $\mathbf{u}_{k-1}$. In view of this, we define a new system with system state $\mathcal{F}_{k-1}$, control input $\mathbf{u}_{k-1}$ and random ``distrurbance" $\mathbf{y}_{k}$, where the state is now perfectly observed.

Before we proceed to the derivation of the DP algorithm, we state two well--known important results, the fundamental lemma of stochastic control and the principle of optimality.

\begin{lem}  [Speyer, Chung \cite{SpeyerChung08}] \label{lem:FLSC} Suppose that the minimum to
\begin{equation*}
\min_{u \in \mathcal{U}} g(x,u)
\end{equation*}

\noindent
exists and $\mathcal{U}$ is a class of functions for which $\mathbb{E}\lbrace g(x,u) \rbrace$ exists. Then,
\begin{equation}
\min_{u(x) \in \mathcal{U}} \mathbb{E} \lbrace g(x,u(x)) \rbrace = \mathbb{E} \lbrace \min_{u(x) \in \mathcal{U}} g(x,u(x)). \rbrace
\end{equation}
\end{lem}

\begin{pri}[Bellman, 1957]
An optimal policy has the property that whatever the initial state and initial decision are, the remaining decisions must constitute an optimal policy with regard to the state resulting from the first decision.
\end{pri}

Theorem \ref{thm:DP_obs_history} gives the DP recursion for computing the optimal control policy for the new system with state $\mathcal{F}_{k-1}$, control input $\mathbf{u}_{k-1}$, and random ``disturbance" $\mathbf{y}_{k}$. 

\begin{thm} \label{thm:DP_obs_history}
For $k = L-1, \dotsc, 1,$ the cost--to--go function $J_{k}(\mathcal{F}_{k-1})$ is related to $J_{k+1}(\mathcal{F}_{k})$ through the recursion
\begin{equation}
\label{eq:DP_obs_history_00}
J_{k}(\mathcal{F}_{k-1}) = \underset{\mathbf{u}_{k-1} \in \mathcal{U}}{\min} \bigg [ \underset{\mathbf{y}_{k}}{\mathbb{E}} \big \lbrace \tr{\big (\mathbf{\Sigma}_{k|k}(\mathbf{y}_{k},\mathbf{u}_{k-1}) \big )} + J_{k+1}(\mathcal{F}_{k-1},\mathbf{y}_{k},\mathbf{u}_{k-1}) \big | \mathcal{F}_{k-1}, \mathbf{u}_{k-1} \big \rbrace \bigg ],
\end{equation}

\noindent
The cost--to--go function for $k = L$ is given by
\begin{equation}
\label{eq:DP_obs_history_01}
J_{L}(\mathcal{F}_{L-1}) = \underset{\mathbf{u}_{L-1} \in \mathcal{U}}{\min} \bigg [ \underset{\mathbf{y}_{L}}{\mathbb{E}} \big \lbrace \tr{ \big ( \mathbf{\Sigma}_{L|L}(\mathbf{y}_{L},\mathbf{u}_{L-1}) \big )} \big | \mathcal{F}_{L-1}, \mathbf{u}_{L-1} \big \rbrace \bigg ].
\end{equation}
\end{thm}

\begin{IEEEproof}
We apply the property of iterated expectation and exploit the conditional independence of the observation--control history to rewrite the optimal cost $J^{*}$ as follows
\begin{IEEEeqnarray}{rCl}
J^{*} & = & \min_{\mathbf{u}_{0},\mathbf{u}_{1},\dotsc,\mathbf{u}_{L-1}} \underset{\mathbf{y}_{0},\mathbf{y}_{1},\dotsc,\mathbf{y}_{L}}{\mathbb{E}} \bigg \lbrace \sum_{k=1}^{L} \tr{\big (\mathbf{\Sigma}_{k|k}(\mathbf{y}_{k},\mathbf{u}_{k-1})\big )} \bigg \rbrace \nonumber \\
& =  &\min_{\mathbf{u}_{0},\mathbf{u}_{1},\dotsc,\mathbf{u}_{L-1}} \mathbb{E} \bigg \lbrace \mathbb{E} \bigg \lbrace \tr \big (\mathbf{\Sigma}_{1|1}(\mathbf{y}_{1},\mathbf{u}_{0}) \big ) + \mathbb{E} \bigg \lbrace \tr \big (\mathbf{\Sigma}_{2|2}(\mathbf{y}_{2},\mathbf{u}_{1}) \big ) + \dotsc +  \mathbb{E} \bigg \lbrace \tr \big ( \mathbf{\Sigma}_{L-1|L-1}(\mathbf{y}_{L-1},\mathbf{u}_{L-2}) \big ) +  \nonumber \\ && \dotsc + \mathbb{E} \bigg \lbrace \tr \big (\mathbf{\Sigma}_{L|L}(\mathbf{y}_{L},\mathbf{u}_{L-1}) \big ) \bigg | \mathcal{F}_{L-1}, \mathbf{u}_{L-1} \bigg \rbrace \bigg | \mathcal{F}_{L-2}, \mathbf{u}_{L-2} \bigg \rbrace
 \bigg | \dotsc \bigg | \mathcal{F}_{1}, \mathbf{u}_{1} \bigg \rbrace
\bigg | \mathcal{F}_{0}, \mathbf{u}_{0} \bigg \rbrace \bigg \rbrace.
\end{IEEEeqnarray}

\noindent
We then use Lemma \ref{lem:FLSC} to interchange the expectation and minimization operations as follows
\begin{IEEEeqnarray}{rCl}
J^{*} & = &\mathbf{E} \bigg \lbrace \min_{\mathbf{u}_{0}} \mathbb{E} \bigg \lbrace \tr \big (\mathbf{\Sigma}_{1|1}(\mathbf{y}_{1},\mathbf{u}_{0}) \big ) + \min_{\mathbf{u}_{1}} \mathbb{E} \bigg \lbrace \tr \big (\mathbf{\Sigma}_{2|2}(\mathbf{y}_{2},\mathbf{u}_{1}) \big ) + \dotsc + \nonumber \\  && \min_{\mathbf{u}_{L-2}} \mathbb{E} \bigg \lbrace \tr \big (\mathbf{\Sigma}_{L-1|L-1}(\mathbf{y}_{L-1},\mathbf{u}_{L-2}) \big ) +\nonumber \\ && \min_{\mathbf{u}_{L-1}} \mathbb{E} \bigg \lbrace \tr \big (\mathbf{\Sigma}_{L|L}(\mathbf{y}_{L},\mathbf{u}_{L-1}) \big ) \bigg | \mathcal{F}_{L-1}, \mathbf{u}_{L-1} \bigg \rbrace  \bigg | \mathcal{F}_{L-2}, \mathbf{u}_{L-2} \bigg \rbrace \bigg | \dotsc \bigg | \mathcal{F}_{1}, \mathbf{u}_{1} \bigg \rbrace \bigg | \mathcal{F}_{0}, \mathbf{u}_{0} \bigg \rbrace  \bigg \rbrace.
\end{IEEEeqnarray}

\noindent
Finally, we employ the principle of optimality to acquire the following recursions
\begin{align*}
J_{L}(\mathcal{F}_{L-1}) &= \underset{\mathbf{u}_{L-1} \in \mathcal{U}}{\min} \bigg [ \underset{\mathbf{y}_{L}}{\mathbb{E}} \big \lbrace \tr{\big (\mathbf{\Sigma}_{L|L}(\mathbf{y}_{L},\mathbf{u}_{L-1})\big )} \big | \mathcal{F}_{L-1}, \mathbf{u}_{L-1} \big \rbrace \bigg ], \\
J_{L-1}(\mathcal{F}_{L-2}) &= \underset{\mathbf{u}_{L-2} \in \mathcal{U}}{\min}  \bigg [ \underset{\mathbf{y}_{L-1}}{\mathbb{E}} \big \lbrace  \tr{\big (\mathbf{\Sigma}_{L|L}(\mathbf{y}_{L-1},\mathbf{u}_{L-2})\big )} + J_{L-1}(\mathcal{F}_{L-2},\mathbf{y}_{L-1},\mathbf{u}_{L-2}) \big | \mathcal{F}_{L-2}, \mathbf{u}_{L-2} \big \rbrace \bigg ], \\
& ~ \thinspace \vdots \\
J_{1}(\mathcal{F}_{0}) &= \underset{\mathbf{u}_{0} \in \mathcal{U}}{\min}  \bigg [ \underset{\mathbf{y}_{1}}{\mathbb{E}} \big \lbrace  \tr{\big (\mathbf{\Sigma}_{1|1}(\mathbf{y}_{1},\mathbf{u}_{0})\big )} + J_{2}(\mathcal{F}_{0},\mathbf{y}_{1},\mathbf{u}_{0}) \big | \mathcal{F}_{0}, \mathbf{u}_{0}\big \rbrace \bigg ],
\end{align*}

\noindent
where the last step concludes the proof.
\end{IEEEproof}

\subsection{Sufficient Statistic \& New DP Algorithm}

As typical with imperfect state information problems, the DP algorithm is carried out over a state space of expanding dimension since the dimension of the state $\mathcal{F}_{k-1}$ increases at each time step $k-1$ with the addition of a new observation. Thus, we seek a sufficient statistic for control purposes (see \emph{e.g.} \cite{BertsekasDPOC05, SpeyerChung08}). For our problem formulation, we can prove by induction \cite{BertsekasDPOC05} that an appropriate sufficient statistic is the conditional probability distribution $\mathbf{p}_{k|k-1}$, which also corresponds to the one-step predicted estimate of the system state.

\begin{prp}
For our active state tracking problem, the conditional distribution $\mathbf{p}_{k|k-1}$ constitutes a sufficient statistic for control purposes.
\end{prp}

\noindent
In one time step, the evolution of this sufficient statistic follows Bayes' rule and is characterized by the following recursive formula
\begin{equation}\label{eq:update_rule_vf}
\mathbf{p}_{k+1|k} = \frac{\mathbf{P} \mathbf{r}(\mathbf{y}_{k},\mathbf{u}_{k-1}) \mathbf{p}_{k|k-1}}{\mathbf{1}_{n}^{T} \mathbf{r}(\mathbf{y}_{k},\mathbf{u}_{k-1}) \mathbf{p}_{k|k-1}},
\end{equation}

\noindent
where $\mathbf{r}(\mathbf{y}_{k},\mathbf{u}_{k-1})$ = $\diag{(f(\mathbf{y}_{k}|\mathbf{e}_{1},\mathbf{u}_{k-1}), \dotsc, f(\mathbf{y}_{k}|\mathbf{e}_{n},\mathbf{u}_{k-1}) )}$ is the $n \times n$ diagonal matrix of measurement vector probability density functions and $\mathbf{1}_{n}$ is a column vector consisting of $n$ ones. Finally, Theorem~\ref{thm:DP_ss_bs} gives the DP algorithm in terms of the sufficient statistic $\mathbf{p}_{k|k-1}$.

\begin{thm} \label{thm:DP_ss_bs}
For $k = L-1, \dotsc, 1,$ the \textit{cost-to-go} function $\overline{J}_{k}(\mathbf{p}_{k|k-1})$ is related to $\overline{J}_{k+1}(\mathbf{p}_{k+1|k})$ through the recursion
\begin{equation}
\label{eq:DP_ss_bs_00}
\overline{J}_{k}(\mathbf{p}_{k|k-1}) = \underset{\mathbf{u}_{k-1} \in \mathcal{U}}{\min} \bigg [ \mathbf{p}_{k|k-1}^{T} \mathbf{h}(\mathbf{p}_{k|k-1}, \mathbf{u}_{k-1}) + \int \mathbf{1}_{n}^{T} \mathbf{r}(\mathbf{y},\mathbf{u}_{k-1}) \mathbf{p}_{k|k-1} \overline{J}_{k+1}\bigg ( \frac{\mathbf{P} \mathbf{r}(\mathbf{y}_{k},\mathbf{u}_{k-1}) \mathbf{p}_{k|k-1}}{\mathbf{1}_{n}^{T} \mathbf{r}(\mathbf{y}_{k},\mathbf{u}_{k-1}) \mathbf{p}_{k|k-1}} \bigg ) d \mathbf{y} \bigg ],
\end{equation}

\noindent
where $\mathbf{h}(\mathbf{p}_{k|k-1}, \mathbf{u}_{k-1})$ is a column vector with components $h(\mathbf{e}_{1},\mathbf{p}_{k|k-1},\mathbf{u}_{k-1}), \dotsc, h(\mathbf{e}_{n},\mathbf{p}_{k|k-1},\mathbf{u}_{k-1})$ with $h(\mathbf{e}_{i},\mathbf{p}_{k|k-1},\mathbf{u}_{k-1}) = 1 - \tr{\big( \mathbf{G}_{k}^{T} \mathbf{G}_{k} \mathbf{Q}_{i}^{\mathbf{u}_{k-1}} \big)} -  \norm{\mathbf{p}_{k|k-1} + \mathbf{G}_{k} (\mathbf{m}_{i}^{\mathbf{u}_{k-1}} - \mathbf{y}_{k|k-1})}^2$. The cost-to-go function for $k = L$ is given by
\begin{equation}
\label{eq:DP_ss_bs_01}
\overline{J}_{L}(\mathbf{p}_{L|L-1}) = \underset{\mathbf{u}_{L-1} \in \mathcal{U}}{\min} \bigg [ \mathbf{p}_{L|L-1}^{T} \mathbf{h}(\mathbf{p}_{L|L-1},\mathbf{u}_{L-1}) \bigg ].
\end{equation}
\end{thm}

\begin{IEEEproof}
For proof, see Appendix \ref{app:pfThm_DP_ss_bs}.
\end{IEEEproof}

Determining the desired control policy via the recursions in Theorem~\ref{thm:DP_ss_bs} results in high computational complexity. Specifically, as with traditional POMDPs, the predicted belief state $\mathbf{p}_{k|k-1}$ is uncountably infinite \cite{BertsekasDPOC05}. Furthermore, the control input definition suggests that the control space size can be exponentially large, while determining the expected future cost is challenging since it requires, in the worst-case, an $N$--dimensional integration for a measurement vector of length $N$. Finally, contrary to standard POMDP problems \cite{BertsekasDPOC05}, the term $\mathbf{p}_{k|k-1}^{T} \mathbf{h}(\mathbf{p}_{k|k-1},\mathbf{u}_{k-1})$ is a \underline{non--linear} function of the predicted belief state $\mathbf{p}_{k|k-1}$ and thus, existing efficient techniques such as \cite{LovejoyAOR91} cannot be directly employed. Still, for small problem sizes, an approximate solution via numerical computation is feasible and can reveal structural characteristics of the optimal solution. Extending techniques from \cite{KrishnamurthyTSP02, KrishnamurthyTSP07}, where non--linear POMDPs have previously appeared can lead to suboptimal but less computational intensive algorithms for deriving the desired control policy. This is out of the scope of the current work and is part of our future research agenda. 

\section{Smoothing Estimators}\label{sec:MMSESSSE}

In this section, we develop suboptimal MMSE smoother formulae for estimating the discrete-time, finite-state Markov chain at each time step. Our goal is to obtain more refined system state estimates given the availability of \underline{both} past and future observations and control inputs. We seek recursive formulae for $\mathbf{p}_{k|s}, s > k$.

 Exploiting the theory introduced in \cite{SegallIT76REDTPP, SegallIT76SPET}, we begin by defining two sequences, similar to the ones in (\ref{eq:estimate_innov_eqn}) and (\ref{eq:obs_innov_eqn}), as follows
\begin{IEEEeqnarray}{rCl}
 \gamma_{s} & \doteq & \mathbf{p}_{k|s} - \mathbf{p}_{k|s-1} = \mathbb{E} \lbrace \mathbf{x}_{k} | \mathcal{F}_{s} \rbrace - \mathbb{E} \lbrace \mathbf{x}_{k} | \mathcal{F}_{s-1} \rbrace, \label{eq:smooth_innov_eqn}\\
 \zeta_{s} & \doteq & \mathbf{y}_{s} - \mathbf{y}_{s|s-1} = \mathbf{y}_{s} - \mathbb{E} \lbrace \mathbf{y}_{s} | \mathcal{F}_{s-1} \rbrace \label{eq:smooth_obs_innov_eqn},
 \end{IEEEeqnarray}

\noindent
 which we can easily prove that are $\lbrace \mathcal{F} \rbrace$--MD sequences. Therefore, the MD representation theorem allows us to write $\lbrace \gamma \rbrace$ in terms of the innovations $\lbrace \zeta \rbrace$ as
\begin{equation}
\gamma_{s} = \mathbf{C}_{s} \zeta_{s}
\end{equation}

\noindent
and $\mathbf{C}_{s}$ can be determined as in (\ref{eq:gain_eqn}) from
\begin{equation}\label{eq:gain_eqn_smooth}
\mathbf{C}_{s} = \mathbb{E} \lbrace \gamma_{s} \zeta_{s}^{T} | \mathcal{F}_{s} \rbrace \big [ \mathbb{E} \lbrace \zeta_{s} \zeta_{s}^{T} | \mathcal{F}_{s} \rbrace \big ]^{-1}.
\end{equation}

\noindent
Once more, the gain sequence $\lbrace C_{s} \rbrace$ is not $\lbrace \mathcal{F} \rbrace$--predictable and thus, to determine a recursive solution, we impose recursivity as a design constraint and use the following approximation
\begin{equation}\label{eq:gain_eqn_smooth_approx}
\mathbf{C}_{s} \approx \mathbb{E} \lbrace \gamma_{s} \zeta_{s}^{T} | \mathcal{F}_{s-1} \rbrace \big [ \mathbb{E} \lbrace \zeta_{s} \zeta_{s}^{T} | \mathcal{F}_{s-1} \rbrace \big ]^{-1}.
\end{equation}

\noindent
Theorem \ref{thm:Smoother_estimate} states the general, finite-dimensional expression for the proposed suboptimal MMSE smoother for the Markov chain system state. 

\begin{thm}\label{thm:Smoother_estimate}
 The R--stage, smoothed estimator of $\mathbf{x}_{k}$, denoted by $\mathbf{p}_{k|R}$ with $R \geqslant k+1, k \geqslant 0$, is given by the expression
\begin{equation}\label{eq:smoother_formula_final}
\mathbf{p}_{k|R} = \mathbf{p}_{k|k} + \sum_{s = k+1}^{R} \mathbf{C}_{s} (\mathbf{y}_{s} - \mathbf{y}_{s|s-1})
\end{equation}

\noindent
with
\begin{equation}\label{eq:smoother_gain_final}
\mathbf{C}_{s} = (\mathbf{\Theta}_{k,s} - \mathbf{p}_{k|s-1} \mathbf{p}^{T}_{s|s-1} ) \mathcal{M}^{T}(\mathbf{u}_{s-1}) (\mathcal{M}(\mathbf{u}_{s-1}) \mathbf{\Sigma}_{s|s-1} \mathcal{M}^{T}(\mathbf{u}_{s-1}) + \widetilde{\mathbf{Q}}_{s})^{-1},
\end{equation}

\noindent
where
\begin{IEEEeqnarray}{rCl}
\mathbf{\Theta}_{k,s}  & = &  \mathbb{E} \lbrace \mathbf{x}_{k} \mathbf{x}_{s-1}^{T} | \mathcal{F}_{s-1} \rbrace \mathbf{P}^{T}, \\
\mathbb{E} \lbrace \mathbf{x}_{k} \mathbf{x}_{s-1}^{T} | \mathcal{F}_{s-1} \rbrace & = & \frac{\mathbf{\Theta}_{k,s-1} \mathbf{r}(\mathbf{y}_{s-1},\mathbf{u}_{s-2})}{\mathbf{1}_{n}^{T} [\mathbf{\Theta}_{k,s-1} \mathbf{r}(\mathbf{y}_{s-1},\mathbf{u}_{s-2})] \mathbf{1}_{n}},
\end{IEEEeqnarray}

\noindent
with $\mathbf{1}_{n}$ denoting the $n \times 1$ vector of ones, $\mathbf{r}(\mathbf{y}_{k},\mathbf{u}_{k-1})$ = $\diag{(f(\mathbf{y}_{k}|\mathbf{e}_{1},\mathbf{u}_{k-1}), \dotsc, f(\mathbf{y}_{k}|\mathbf{e}_{n},\mathbf{u}_{k-1}) )}$ the $n \times n$ diagonal matrix of measurement vector probability density functions, $\mathbb{E} \lbrace \mathbf{x}_{0} \mathbf{x}_{0}^{T} | \mathcal{F}_{0} \rbrace = \diag(\mathbf{p}_{0|0})$ and $\widetilde{\mathbf{Q}}_{s} = \sum_{i=1}^{n} p_{s|s-1}^{i} \mathbf{Q}_{i}^{\mathbf{u}_{s-1}}$.
\end{thm}
 
\begin{IEEEproof}
For proof, see Appendix \ref{app:pfThm_Smoother_estimate}.
\end{IEEEproof}

The MSE performance of the smoother in (\ref{eq:smoother_formula_final}) can be calculated similarly to the MSE performance of the filter and is characterized by the \textit{conditional smoothing error covariance matrix}
\begin{equation}
\mathbf{\Sigma}_{k|R} \doteq \mathbb{E} \lbrace (\mathbf{x}_{k} - \mathbf{p}_{k|R}) (\mathbf{x}_{k} - \mathbf{p}_{k|R})^{T} | \mathcal{F}_{R} \rbrace = \diag (\mathbf{p}_{k|R}) - \mathbf{p}_{k|R} \mathbf{p}_{k|R}^{T},~R \geqslant k+1,~k \geqslant 0.
\end{equation}

As evident from Theorem \ref{thm:Smoother_estimate}, the gain matrix $\mathbf{C}_{s}$ depends \textit{non-linearly} on the observations, as it is the case with the Kalman--like filter. Comparing our smoothed estimator with the corresponding Kalman smoother \cite{ChenTR03}, we observe that in both cases filtered estimates are required to obtain smoothed estimates and the smoothers gains do not depend on conditional smoothing error covariance matrices. Furthermore, as with the standard Kalman smoother, the Kalman--like filter's gain is a factor of the smoother gain since
\begin{equation}
\mathbf{C}_{s} = (\mathbf{\Theta}_{k,s} - \mathbf{p}_{k|s-1} \mathbf{p}^{T}_{s|s-1} ) \mathbf{\Sigma}_{s|s-1}^{-1} \mathbf{G}_{s},
\end{equation} 

\noindent
and this allows us to rewrite the smoother in (\ref{eq:smoother_formula_final}) as follows
\begin{equation}
\mathbf{p}_{k|R} = \mathbf{p}_{k|k} + \sum_{s = k+1}^{R} (\mathbf{\Theta}_{k,s} - \mathbf{p}_{k|s-1} \mathbf{p}^{T}_{s|s-1} ) \mathbf{\Sigma}_{s|s-1}^{-1} (\mathbf{p}_{s+1|s} - \mathbf{p}_{s+1|s}).
\end{equation}

There are three well-known types of smoothers in the literature depending on the way observations are processed: \emph{fixed--point}, \emph{fixed--interval}, and \emph{fixed--lag}. The \emph{fixed--point} smoother $\mathbf{p}_{k|R}, R \geqslant k+1$, uses all available information up to and including time step $R$, to improve the estimate of a state at a specific time step. On the contrary, the \emph{fixed--interval} smoother $\mathbf{p}_{k|L} \doteq \mathbb{E} \lbrace \mathbf{x}_{k} | \mathcal{F}_{L} \rbrace , k = 0, 1,  \dotsc, L - 1,$ uses all available information, while the \emph{fixed--lag} smoother $\mathbf{p}_{k|k + \Delta} \doteq \mathbb{E} \lbrace \mathbf{x}_{k} | \mathcal{F}_{k + \Delta} \rbrace , k = 0, 1,  \dotsc,$ uses all information up to and including a fixed interval of time $\Delta$ from the time step of interest. Theorem \ref{thm:Smoother_estimate}, Propositions \ref{prp:fi_smoother} and \ref{prp:fl_smoother} give the expressions for the fixed--point, the fixed--interval and the fixed--lag smoothed estimators, respectively.

\begin{prp}\label{prp:fi_smoother}
The fixed--interval smoothed estimator of $\mathbf{x}_{k}, \mathbf{p}_{k|L},$ is given by the expression
\begin{equation}\label{eq:fi_smoother}
\mathbf{p}_{k|L} = \mathbf{P} \mathbf{p}_{k-1|L} + (\mathbf{I}_{n} - \mathbf{P}) \sum_{s = k}^{L} \mathbf{C}_{s} (\mathbf{y}_{s} - \mathbf{y}_{s|s-1}), ~k = 1, 2, \dotsc , L-1,
\end{equation}

\noindent
where $\mathbf{I}_{n}$ is the $n \times n$ identity matrix, and is initialized by
\begin{equation}
\mathbf{p}_{0|L} = \mathbf{p}_{0|0} + \sum_{s=1}^{L} \mathbf{C}_{s} (\mathbf{y}_{s} - \mathbf{y}_{s|s-1}),
\end{equation}

\noindent
which is obtained from the fixed--point smoothed estimator by setting $k = 0$.
\end{prp}
\begin{IEEEproof}
For proof, see Appendix \ref{app:pf_fi_smoother}.
\end{IEEEproof}

\begin{prp}\label{prp:fl_smoother}
The fixed--lag smoothed estimator of $\mathbf{x}_{k}, \mathbf{p}_{k|k+\Delta},$ is given by the expression
\begin{equation}\label{eq:fl_smoother}
\mathbf{p}_{k|k+\Delta} = \mathbf{P} \mathbf{p}_{k-1|k + \Delta - 1} + \Gamma(k,\Delta) + (\mathbf{I}_{n} - \mathbf{P}) \sum_{s = k+1}^{k+\Delta-1} \mathbf{C}_{s} (\mathbf{y}_{s} - \mathbf{y}_{s|s-1}), ~k = 0, 1, \dotsc,
\end{equation}

\noindent
where $\mathbf{I}_{n}$ is the $n \times n$ identity matrix, $\Gamma(k,\Delta)$ is defined as
\begin{equation}
\Gamma(k,\Delta) \doteq \mathbf{C}_{k+\Delta} (\mathbf{y}_{k+\Delta} - \mathbf{y}_{k+\Delta | k + \Delta - 1}) - \mathbf{p}_{k+1|k} - \mathbf{p}_{k|k-1} + \mathbf{P} \mathbf{p}_{k|k-1},
\end{equation}

\noindent
and the smoother is initialized by
\begin{equation}
\mathbf{p}_{0|\Delta} = \mathbf{p}_{0|0} + \sum_{s=1}^{\Delta} \mathbf{C}_{s} (\mathbf{y}_{s} - \mathbf{y}_{s|s-1}),
\end{equation}

\noindent
which is obtained from the fixed--point smoothed estimator by setting $k = 0$.
\end{prp}

\begin{IEEEproof}
For proof, see Appendix \ref{app:pf_fl_smoother}.
\end{IEEEproof}

\noindent
We underscore that, as in the case of the Kalman--like estimator, we can apply a suitable memoryless (linear or nonlinear) transformation to the smoothed estimates above to obtain valid probability mass functions.

\section{Numerical Example}\label{sec:EX}

In this section, we provide numerical results to illustrate the performance of the proposed framework for a body sensing application \cite{ZoisTSP13}. The goal is to estimate the time-evolving physical activity state of an individual by using information from three biometric sensors: two accelerometers (ACCs) and an electrocardiograph (ECG). We focus on distinguishing between four physical states (\textit{Sit}, \textit{Stand}, \textit{Run}, \textit{Walk}) with transition probability matrix $\mathbf{P}$ of the form
\begin{equation*}
\mathbf{P} = 
\begin{bmatrix}
0.6 & 0.2 & 0 & 0.4 \\
0.1 & 0.4 & 0.1 & 0 \\
0 & 0.1 & 0.3 & 0.3 \\
0.3 & 0.3 & 0.6 & 0.3
\end{bmatrix}.
\end{equation*}

\noindent
The control input is defined as a tuple with each element indicating the requested number of samples from the associated sensor at each time step, while the total requested number of samples does not exceed a budget of $N$ samples. Each sample corresponds to an extracted feature value from the associated biometric signal and here, we focus on three features: 1) the ACC mean from the first ACC ($S_{1}$), 2) the ACC variance from the second ACC ($S_{2}$), and 3) the ECG period from the ECG ($S_{3}$). Based on the problem characteristics and the control input definition, the signal model in (\ref{eq:observation_model}) constitutes an AR(1)--correlated multivariate conditionally Gaussian model with statistics
\begin{IEEEeqnarray}{rCl}
\mathbf{m}_{i}^{ \mathbf{u}_{k-1} } & = & [\mu_{i,S_{1}}^{\mathbf{u}_{k-1}}, \mu_{i,S_{2}}^{\mathbf{u}_{k-1}}, \mu_{i,S_{3}}^{\mathbf{u}_{k-1}}]^T,\\
\mathbf{Q}_{i}^{ \mathbf{u}_{k-1} } & = & \diag (\mathbf{Q}_{i}^{ \mathbf{u}_{k-1} }(S_{1}), \mathbf{Q}_{i}^{ \mathbf{u}_{k-1} }(S_{2}), \mathbf{Q}_{i}^{ \mathbf{u}_{k-1} }(S_{3})), \\
\mathbf{Q}_{i}^{ \mathbf{u}_{k-1} }(S_{l}) & = & \frac{\sigma_{S_{l},i}^{2}}{1-\phi^2} \mathbf{T} + \sigma_{z}^{2} \mathbf{I},
\end{IEEEeqnarray}

\noindent
where $i$ indicates physical state $\mathbf{e}_{i}$, $S_{l}$ denotes sensor $l$, $\mu_{i,S_{l}}^{\mathbf{u}_{k-1}}$ is of size $N_{l}^{ \mathbf{u}_{k-1} } \times 1$, $\mathbf{T}$ is a Toeplitz matrix with first row/column $[1,\phi,\phi^2,\dotsc,\phi^{N_{l}^{\mathbf{u}_{k-1}} -1}]$, $\mathbf{I}$ is the $N_{l}^{ \mathbf{u}_{k-1} } \times N_{l}^{ \mathbf{u}_{k-1} }$ identity matrix, $N_{l}^{ \mathbf{u}_{k-1} }$ indicates the requested number of samples from sensor $S_{l}$, $\phi$ is the parameter of the AR(1) model and $\sigma_{z}^{2}$ accounts for sensing and communication noise. For more information about the model, we refer the interested reader to \cite{ZoisTSP13}.

\begin{figure}[h!]
\centering
{\includegraphics[width=0.6\columnwidth]{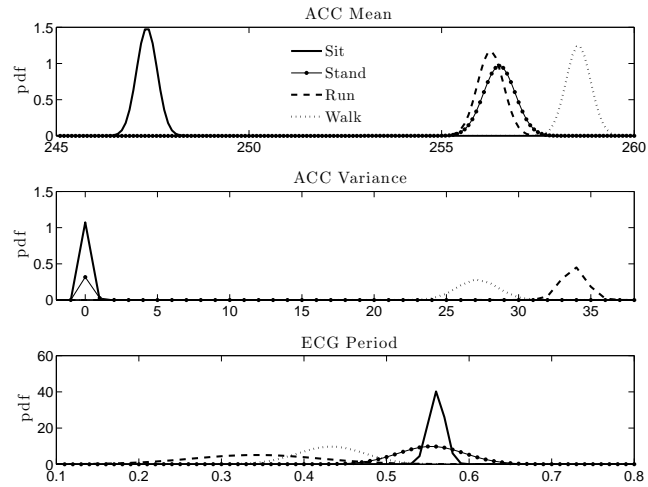}}
\caption{Gaussian distributions associated with each of the four activity states for the ACC mean, ACC variance and ECG period features, respectively. The plots indicate that a combination of samples from the ACC mean and the ACC variance can help us discriminate between the physical activities of interest. On the other hand, the ECG Period is not very informative.}
\label{fig:4Hypotheses}
\end{figure}

Our numerical simulations are based on the above model and driven by \underline{real} data collected by a prototype body sensing network, the KNOWME network \cite{ThatteTSP11}. Data collection was conducted in the lab and consisted of three to four sessions, where twelve subjects performed eight physical activities. A detailed description of the data collection protocol and subject characteristics can be found in \cite{ThatteTSP11}. Herein, to showcase our framework's performance, we focus on distinguishing between four activities for a single individual with signal model distributions shown in Fig.~\ref{fig:4Hypotheses}. We have assumed  for a budget of $N = 2$ samples, $\phi = 0.25$ and $\sigma_{z}^{2} = 2$. Even though the number of samples are few, patterns still emerge. We underscore that our methods are directly applicable to multiple sensors and physical states as well as larger budget of samples.

We begin by numerically establishing the suboptimalilty of our proposed Kalman--like estimator. Specifically, we test its optimality by numerically comparing its performance with the performance of the optimal MMSE estimator. More precisely, for our system model of interest, the optimal MMSE estimate can be recursively determined via Bayes' rule as follows
\begin{equation}\label{eq:optimal_BR_estimator}
\mathbf{p}_{k|k} = \frac{\mathbf{r}(\mathbf{y}_{k},\mathbf{u}_{k-1}) \mathbf{P} \mathbf{p}_{k-1|k-1}}{\mathbf{1}_{n}^{T} \mathbf{r}(\mathbf{y}_{k},\mathbf{u}_{k-1}) \mathbf{P}\mathbf{p}_{k-1|k-1}}.
\end{equation}

\begin{figure}[h!]
\centering
{\includegraphics[width=0.6\columnwidth]{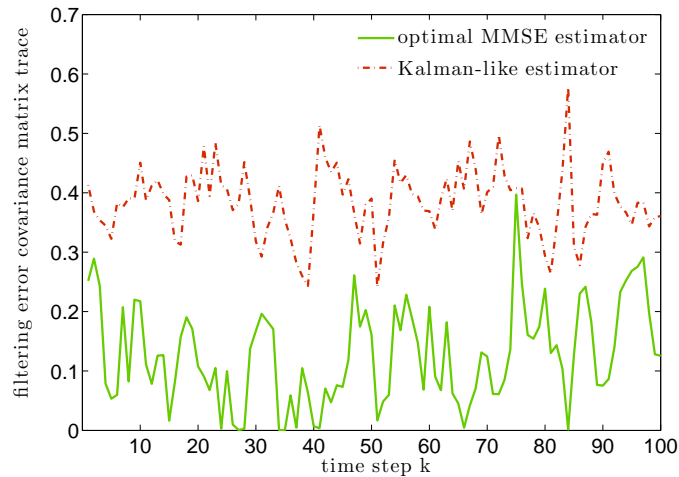}}
\caption{Average MSE performance of optimal MMSE estimator and Kalman--like estimator.}
\label{fig:comparison_MSE_Bayes_vs_filter}
\end{figure}

\noindent
In Fig.~\ref{fig:comparison_MSE_Bayes_vs_filter}, the MSE performance (trace of filtering error covariance matrix) of the Kalman--like estimator in $(\ref{eq:filter_eqn})$ and the optimal MMSE estimator in $(\ref{eq:optimal_BR_estimator})$ are shown. Comparing the MSE performance of the two estimators, we observe that the proposed Kalman--like estimator achieves higher MSE compared to the optimal MMSE estimator. This fact implies that the former estimator is suboptimal in the sense that it results in higher MSE on average. In addition, using an MAP rule on top of the two estimators results in $87\%$ and $92\%$ detection accuracy for the Kalman--like filter and the optimal MMSE estimator, respectively. This fact reinforces our belief that the proposed estimator must be suboptimal.

\begin{figure}[h!]
\centering
{\includegraphics[width=0.6\columnwidth]{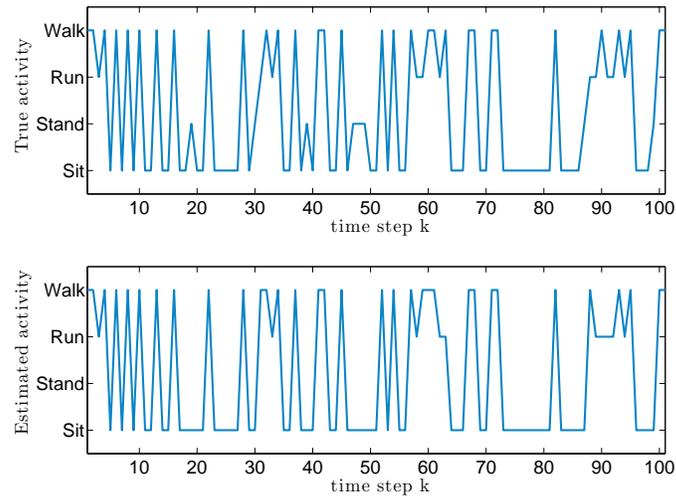}}
\caption{Tracking performance of the proposed framework: the upper plot shows the individual's true activity, while the lower plot the estimated activity.}
\label{fig:tracking_performance}
\end{figure}

Next, in Fig.~\ref{fig:tracking_performance}, we present the tracking performance of the proposed framework (Kalman--like estimator and optimal policy design) by showing the true and estimated state sequences. The output of our system is an estimate of the belief state and we estimate our activity state via a MAP rule. We observe that the proposed framework tracks significantly well the underlying, time-evolving activity state even though the total number of samples used are few. Furthermore, we note that the \textit{Stand} state is usually not detected since according to the stationary distribution of the Markov chain, it corresponds to an ephemeral state. Modifications of the tracking cost similar to the ones presented in \cite{ZoisTSP13} can be employed to detect ephemeral states.

Table \ref{tb:detection_accuracy} summarizes the detection accuracy achieved by employing different control policies. The different control policies are: 1) always select one sample from ACC Mean (strategy A), 2) always select one sample from ACC Variance (strategy B), 3) always select one sample from ECG period (strategy $\Gamma$), and 4) optimal sensor selection policy. We find that selecting a control strategy independent of the estimated belief (strategies A, B, $\Gamma$) does not benefit the detection accuracy. Similar is the case of using only one sample from one of the available sensors unless the selected sensor can discriminate easily between all the states. Furthermore, fusing samples from sensors of different capabilities, as done by the optimal control policy, can boost detection performance significantly. Finally, we expect that for larger values of the total number $N$ of available samples, the detection accuracy would be even higher.

At this point, we wish to comment on the form of the optimal control policy. The optimal control policy consists of three types of control inputs: 1) ACC mean -- 2 samples, 2) ACC mean -- 1 sample and ACC variance -- 1 sample, and 3) ACC mean -- 2 samples. The first type of control input is selected for most of the predicted belief states and this is due to the fact that it can discriminate between the more likely states \emph{i.e.} \textit{Sit}, \textit{Run}, \textit{Walk}. The second and third types of control input are primarily selected for detecting the least likely state, \textit{Stand}. Specifically, when the Sit state has low probability ($\leqslant 0.5$), the  second control input is selected since one sample from each of the informative sensors can help us discriminate \textit{Stand} from the rest of the states. However, when the \textit{Run} and \textit{Walk} states have zero probability, samples from ACC mean are enough to detect \textit{Stand}, as verified by Fig.~\ref{fig:4Hypotheses}.

\begin{figure}[h!]
   \centering
   \includegraphics[width=0.6\linewidth]{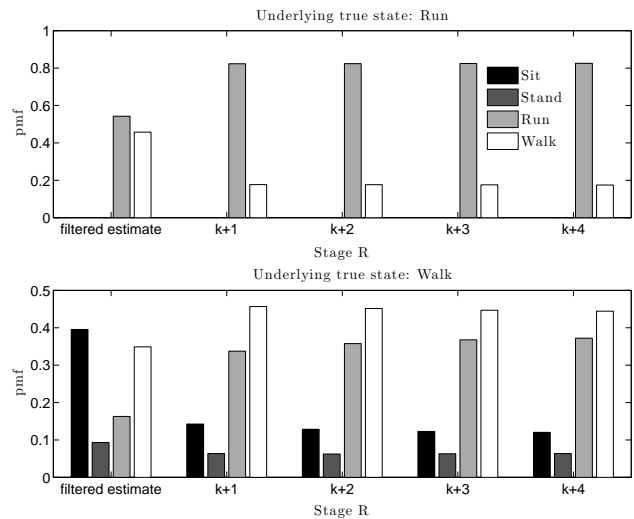}
   \caption{Exemplary effect of stage $R$ on the smoothed state estimates (pmfs). The initial filtered estimate is also given for comparison.}
   \label{fig:smoothing_R_effect}
 \end{figure}

Finally, Table \ref{tab:detection_accuracy} summarizes the detection accuracy of filtering and smoothing operations. We observe that as expected, smoothing enhances detection accuracy. However, also expected, the smoothing performance saturates as the stage $R$ increases. We underscore that different Markov chains and/or signal model statistics would result in different smoothing performance improvements. In Fig.~\ref{fig:smoothing_R_effect}, we present an example of the effect of increasing the smoother's stage on the pmf over the underlying state. We observe that future information can enhance or overturn our belief with respect to the true system state unveiling its true value. As $R$ increases, our belief stabilizes, which is also is indicated by the results in Table \ref{tab:detection_accuracy}. Finally, even though the detection accuracy does not improve significantly as $R$ increases, the associated MSE does, as supported by the results in Table \ref{tab:detection_accuracy}.

\section{Conclusions \& Future Work}\label{sec:CON}

In this work, we addressed the active state tracking problem for a discrete--time, finite--state Markov chain observed via conditionally Gaussian measurements. We proposed a unified framework that combines MMSE state estimation (prediction, filtering and smoothing) and control policy design. Following an innovations method, we derived a non--linear Kalman--like estimator for the Markov chain system state, which is formally similar to the classical KF. We also derived a stochastic dynamic programming algorithm to determine the optimal control policy with the cost functional being the filters' MSE performance. To enhance state estimation performance, we derived recursive formulae for the three fundamental smoothing types (fixed--point, fixed--lag, fixed--interval). Finally, we verified the successfulness of our proposed framework on a body sensing application using real data collected from a prototype body sensing network. Our results differ from prior work in that we jointly consider time-varying systems, discrete states and  active control over measurements. We believe that our framework is widely applicable to a broad spectrum of active classification applications including sensor management for object classification and control, radar scheduling and estimation of sparse signals.

At this stage, approximate optimal control policies were determined by numerically solving the DP equation. Our current efforts involve the structural characterization of the optimal control policy, based on which computationally efficient control strategies will be proposed. Future work will focus on considering sensing usage costs and addressing applications admitting our framework.

\begin{appendix}

\subsection{Proof of Theorem \ref{thm:MMSE_estimate}}\label{app:pfThm_MMSE_estimate}

Having defined the estimate and observations innovation sequences as in (\ref{eq:estimate_innov_eqn}) and (\ref{eq:obs_innov_eqn}), we apply (\ref{eq:Kalman_like_eqn}) to get the desired recursive filter equation. To this end, we need to determine a recursive form that relates $\mathbf{p}_{k|k-1}$ to $\mathbf{p}_{k-1|k-1}$ and explicit formulas for $\mathbf{y}_{k|k-1}$ and $\mathbf{G}_{k}$.

The expected value of $\mathbf{x}_{k}$ conditioned on the observation-control history $\mathcal{F}_{k-1}$ can be determined as follows
\begin{IEEEeqnarray}{rCl}
\mathbf{p}_{k|k-1} & = & \mathbb{E} \lbrace \mathbf{x}_{k} | \mathcal{F}_{k-1} \rbrace = \mathbf{P} \mathbb{E} \lbrace \mathbf{x}_{k-1} | \mathcal{F}_{k-1} \rbrace + \mathbb{E} \lbrace \mathbf{w}_{k} | \mathcal{F}_{k-1} \rbrace \label{eq:state_predictor_deriv00} \\ & \overset{(a)}= & \mathbf{P} \mathbf{p}_{k-1|k-1} + \mathbb{E} \lbrace \mathbb{E} \lbrace \mathbf{w}_{k} | \mathcal{B}_{k-1} \rbrace | \mathcal{F}_{k-1} \rbrace \label{eq:state_predictor_deriv01}\\ & \overset{(b)}= & \mathbf{P} \mathbf{p}_{k-1|k-1} \label{eq:state_predictor},
\end{IEEEeqnarray}

\noindent
where we have exploited $(a)$ the  law of iterated expectations for $\sigma$--algebras and $(b)$ the fact that $\mathbf{w}_{k}$ is a $\lbrace \mathcal{B} \rbrace$--MD sequence. For the the initial condition, we have $\mathbf{p}_{0|-1} \doteq \mathbb{E} \lbrace \mathbf{x}_{0} \rbrace = \pi$, where $\pi$ is the initial distribution over the system states. Similarly, the expected value of the process $\lbrace \mathbf{y}_{k} \rbrace$ conditioned on the observation-control history $\mathcal{F}_{k-1}$ can be determined as follows
\begin{IEEEeqnarray}{rCl}
\mathbf{y}_{k|k-1} & = & \mathbb{E} \lbrace  \mathbf{y}_{k} | \mathcal{F}_{k-1} \rbrace = \mathbb{E} \lbrace \mathcal{M}(\mathbf{u}_{k-1})\mathbf{x}_{k} + \mathbf{v}_{k} | \mathcal{F}_{k-1} \rbrace \\ & \overset{(a)}= & \mathcal{M}(\mathbf{u}_{k-1}) \mathbb{E} \lbrace \mathbf{x}_{k} | \mathcal{F}_{k-1} \rbrace + \mathbb{E} \lbrace \mathbf{v}_{k} | \mathcal{F}_{k-1} \rbrace \overset{(b)}= \mathcal{M}(\mathbf{u}_{k-1}) \mathbf{p}_{k|k-1} \label{eq:obs_predictor},
\end{IEEEeqnarray}

\noindent
where we have exploited that $(a)$ $\mathbf{u}_{k-1} = \eta_{k-1}(\mathcal{F}_{k-1})$ and $(b)$ $\mathbf{v}_{k}$ is a $\lbrace \mathcal{B}^{-} \rbrace$--MD sequence.

At this point, we can specify each of the terms that comprise the filter gain in (\ref{eq:gain_eqn}). Specifically, for $\mathbb{E} \lbrace \lambda_{k} \lambda_{k}^{T} | \mathcal{F}_{k-1} \rbrace$, we have
\begin{IEEEeqnarray}{rCl}
\mathbb{E} \lbrace \lambda_{k} \lambda_{k}^{T} | \mathcal{F}_{k-1} \rbrace = \mathbb{E} \lbrace (\mathbf{y}_{k} - \mathbf{y}_{k|k-1})(\mathbf{y}_{k} - \mathbf{y}_{k|k-1})^{T} | \mathcal{F}_{k-1}\rbrace = \mathbb{E} \lbrace \mathbf{y}_{k} \mathbf{y}_{k}^{T} | \mathcal{F}_{k-1} \rbrace - \mathbf{y}_{k|k-1} \mathbf{y}_{k|k-1}^{T} \label{eq:variance_eqn}.
\end{IEEEeqnarray}

\noindent
In order to determine the exact form of $\mathbb{E} \lbrace \mathbf{y}_{k} \mathbf{y}_{k}^{T} | \mathcal{F}_{k-1}\rbrace$, we first need to determine $p(\mathbf{y}_{k}|\mathcal{F}_{k-1})$. Thus, we work as follows
\begin{IEEEeqnarray}{rCl}
p(\mathbf{y}_{k}|\mathcal{F}_{k-1}) & = & p(\mathbf{y}_{k} | \mathbf{y}_{0}, \dotsc, \mathbf{y}_{k-1}, \mathbf{u}_{0}, \dotsc, \mathbf{u}_{k-2}) \overset{(a)}= p(\mathbf{y}_{k} | \mathbf{y}_{0}, \dotsc, \mathbf{y}_{k-1}, \mathbf{u}_{0}, \dotsc, \mathbf{u}_{k-1}) 
\\ & = & \sum_{i=1}^{n} P(\mathbf{x}_{k} = \mathbf{e}_{i} | \mathbf{y}_{0}, \dotsc, \mathbf{y}_{k-1}, \mathbf{u}_{0}, \dotsc, \mathbf{u}_{k-1}) p(\mathbf{y}_{k} | \mathbf{x}_{k} = \mathbf{e}_{i}, \mathbf{u}_{k-1}) \\ & \overset{(b)}=& \sum_{i=1}^{n} P(\mathbf{x}_{k} = \mathbf{e}_{i} | \mathcal{F}_{k-1}) f(\mathbf{y}_{k} | \mathbf{e}_{i}, \mathbf{u}_{k-1}) = \sum_{i=1}^{n} p_{k|k-1}^{i} f(\mathbf{y}_{k} | \mathbf{e}_{i}, \mathbf{u}_{k-1}), \label{eq:cond_pdf}
\end{IEEEeqnarray}

\noindent
where for $(a), (b),$ we have exploited that  $\mathbf{u}_{k-1} = \eta_{k-1}(\mathcal{F}_{k-1})$. The last result implies that
\begin{IEEEeqnarray}{rCl}
\mathbb{E} \lbrace \mathbf{y}_{k} \mathbf{y}_{k}^{T} | \mathcal{F}_{k-1}\rbrace & = & \int \mathbf{y} \mathbf{y}^{T} p(\mathbf{y}|\mathcal{F}_{k-1}) d\mathbf{y} = \sum_{i=1}^{n} p_{k|k-1}^{i} \int \mathbf{y} \mathbf{y}^{T} f(\mathbf{y} | \mathbf{e}_{i}, \mathbf{u}_{k-1}) d\mathbf{y} \\ & = & \sum_{i=1}^{n} p_{k|k-1}^{i} \big [ \mathbf{Q}_{i}^{\mathbf{u}_{k-1}} + \mathbf{m}_{i}^{\mathbf{u}_{k-1}} {(\mathbf{m}_{i}^{\mathbf{u}_{k-1}})^{T}}\big ],
\end{IEEEeqnarray}

\noindent
and substituting back to (\ref{eq:variance_eqn}), and performing some manipulations, we get
\begin{IEEEeqnarray}{rCl}
\mathbb{E} \lbrace \lambda_{k} \lambda_{k}^{T} | \mathcal{F}_{k-1} \rbrace & = & \sum_{i=1}^{n} p_{k|k-1}^{i} \big [ \mathbf{Q}_{i}^{\mathbf{u}_{k-1}} + \mathbf{m}_{i}^{\mathbf{u}_{k-1}} {(\mathbf{m}_{i}^{\mathbf{u}_{k-1}})^{T}}\big ] - \mathbf{y}_{k|k-1} \mathbf{y}_{k|k-1}^{T} \\
& = & \sum_{i=1}^{n} p_{k|k-1}^{i} \mathbf{Q}_{i}^{\mathbf{u}_{k-1}} + \mathcal{M}(\mathbf{u}_{k-1}) \diag{(\mathbf{p}_{k|k-1})} \mathcal{M}^{T}(\mathbf{u}_{k-1}) \nonumber \\ && - \mathcal{M}(\mathbf{u}_{k-1})  \mathbf{p}_{k|k-1} \mathbf{p}_{k|k-1}^{T} \mathcal{M}^{T}(\mathbf{u}_{k-1}) \\
& = & \widetilde{\mathbf{Q}}_{k} + \mathcal{M}(\mathbf{u}_{k-1}) (\diag{(\mathbf{p}_{k|k-1})} - \mathbf{p}_{k|k-1} \mathbf{p}_{k|k-1}^{T}) \mathcal{M}^{T}(\mathbf{u}_{k-1}) \\
& = & \widetilde{\mathbf{Q}}_{k} + \mathcal{M}(\mathbf{u}_{k-1}) \mathbf{\Sigma}_{k|k-1} \mathcal{M}^{T}(\mathbf{u}_{k-1}), \label{eq:second_term_gain}
\end{IEEEeqnarray}

\noindent
where $\widetilde{\mathbf{Q}}_{k} = \sum_{i=1}^{n} p_{k|k-1}^{i} \mathbf{Q}_{i}^{\mathbf{u}_{k-1}}$ and we have also used the definition of the conditional prediction error covariance matrix in (\ref{eq:ccmfe}).
Next, we derive the term $\mathbb{E} \lbrace \mu_{k} \lambda_{k}^{T} | \mathcal{F}_{k-1} \rbrace$. Specifically:
\begin{equation}\label{eq:covariance_eqn}
\mathbb{E} \lbrace \mu_{k} \lambda_{k}^{T} | \mathcal{F}_{k-1} \rbrace = \mathbb{E} \lbrace \mathbf{p}_{k|k} \lambda_{k}^T | \mathcal{F}_{k-1} \rbrace - \mathbb{E} \lbrace \mathbf{p}_{k|k-1} \lambda_{k}^T | \mathcal{F}_{k-1} \rbrace.
\end{equation}

\noindent
The first term of (\ref{eq:covariance_eqn}) is determined as follows
\begin{IEEEeqnarray}{rCl}
\mathbb{E} \lbrace \mathbf{p}_{k|k} \lambda_{k}^T | \mathcal{F}_{k-1} \rbrace & = & \mathbb{E} \lbrace \mathbb{E} \lbrace \mathbf{x}_{k} | \mathcal{F}_{k} \rbrace \lambda_{k}^T | \mathcal{F}_{k-1} \rbrace =  \mathbb{E} \lbrace \mathbb{E} \lbrace \mathbf{x}_{k} \lambda_{k}^T | \mathcal{F}_{k} \rbrace | \mathcal{F}_{k-1} \rbrace \label{eq:first_term_00} \\ & = & \mathbb{E} \lbrace \mathbf{x}_{k} (\mathbf{y}_{k} - \mathbf{y}_{k|k-1})^T | \mathcal{F}_{k-1} \rbrace = \mathbb{E} \lbrace \mathbf{x}_{k} \mathbf{y}_{k}^{T} | \mathcal{F}_{k-1} \rbrace - \mathbf{p}_{k|k-1}\mathbf{y}_{k|k-1}^T,
\end{IEEEeqnarray}

\noindent
where we have exploited the MD property of $\lambda_{k}$. The term $\mathbb{E} \lbrace \mathbf{x}_{k} \mathbf{y}_{k}^{T} | \mathcal{F}_{k-1} \rbrace$ can be determined as follows
\begin{IEEEeqnarray}{rCl}
\mathbb{E} \lbrace \mathbf{x}_{k} \mathbf{y}_{k}^{T} | \mathcal{F}_{k-1} \rbrace 
& = & \mathbb{E} \lbrace \mathbf{x}_{k} \mathbf{x}_{k}^{T} \mathcal{M}^{T}(\mathbf{u}_{k-1}) | \mathcal{F}_{k-1} \rbrace + \mathbb{E} \lbrace \mathbf{x}_{k}\mathbf{v}_{k}^{T} | \mathcal{F}_{k-1} \rbrace  \\
& = & \sum_{i=1}^{n} \mathbf{e}_{i} \mathbf{e}_{i}^{T} P(\mathbf{x}_{k} = \mathbf{e}_{i} | \mathcal{F}_{k-1}) \mathcal{M}^{T}(\mathbf{u}_{k-1})  + \mathbb{E} \lbrace \mathbb{E} \lbrace \mathbf{x}_{k} \mathbf{v}_{k}^{T} | \mathcal{B}_{k}^{-} \rbrace | \mathcal{F}_{k-1} \rbrace \label{eq:intermediate_term_00}\\
& = & \diag{(\mathbf{p}_{k|k-1})}\mathcal{M}^{T}(\mathbf{u}_{k-1})  + \mathbb{E} \lbrace \mathbf{x}_{k} \mathbb{E} \lbrace \mathbf{v}_{k}^{T} | \mathcal{B}_{k}^{-} \rbrace | \mathcal{F}_{k-1} \rbrace \label{eq:intermediate_term_01}\\
& = & \diag{(\mathbf{p}_{k|k-1})}\mathcal{M}^{T}(\mathbf{u}_{k-1}),
\end{IEEEeqnarray}

\noindent
where we have used the facts that $\mathbf{u}_{k-1} = \eta_{k-1}(\mathcal{F}_{k-1})$, $\mathbf{x}_{k} \in \mathcal{B}_{k}^{-}$ by definition and the MD property of  $\mathbf{v}_{k}$. The second term of (\ref{eq:covariance_eqn}) is determined as follows
\begin{equation}
\mathbb{E} \lbrace \mathbf{p}_{k|k-1} \lambda_{k}^T | \mathcal{F}_{k-1} \rbrace = \mathbf{p}_{k|k-1} \mathbb{E} \lbrace \lambda_{k}^{T} | \mathcal{F}_{k-1} \rbrace = 0,
\end{equation}

\noindent
where the last equality holds since $\lambda_{k}$ is a $\lbrace\mathcal{F}\rbrace$--MD sequence. Combing the above results, we have
\begin{IEEEeqnarray}{rCl}
\mathbb{E} \lbrace \mu_{k} \lambda_{k}^{T} | \mathcal{F}_{k-1} \rbrace & = & \diag{(\mathbf{p}_{k|k-1})}\mathcal{M}^{T}(\mathbf{u}_{k-1}) - \mathbf{p}_{k|k-1}\mathbf{y}_{k|k-1}^T \\ & = & (\diag{(\mathbf{p}_{k|k-1})} - \mathbf{p}_{k|k-1} \mathbf{p}_{k|k-1}^{T})\mathcal{M}^{T}(\mathbf{u}_{k-1}) = \mathbf{\Sigma}_{k|k-1} \mathcal{M}^{T}(\mathbf{u}_{k-1}),
\end{IEEEeqnarray}

\noindent
and the gain $\mathbf{G}_{k}$ takes the following form
\begin{equation}\label{eq:new_gain_eqn}
\mathbf{G}_{k} =  \mathbf{\Sigma}_{k|k-1} \mathcal{M}^{T}(\mathbf{u}_{k-1})(\mathcal{M}(\mathbf{u}_{k-1}) \mathbf{\Sigma}_{k|k-1}\mathcal{M}^{T}(\mathbf{u}_{k-1}) + \widetilde{\mathbf{Q}}_{k})^{-1}.
\end{equation}

\noindent
Therefore, using (\ref{eq:estimate_innov_eqn}) and (\ref{eq:obs_innov_eqn}), (\ref{eq:Kalman_like_eqn}) can be rewritten as:
\begin{equation*}
\mathbf{p}_{k|k} = \mathbf{p}_{k|k-1} + \mathbf{G}_{k} [\mathbf{y}_{k} - \mathbf{y}_{k|k-1} ],~ k \geqslant 0,
\end{equation*}

\noindent
and together with (\ref{eq:state_predictor}), (\ref{eq:obs_predictor})  and (\ref{eq:new_gain_eqn}) constitute a recursive exact algorithm for the computation of the belief state defined in (\ref{eq:APP_vector}).

\subsection{Proof of Theorem \ref{thm:DP_ss_bs}}\label{app:pfThm_DP_ss_bs}

Before we proceed with the proof of Theorem \ref{thm:DP_ss_bs}, we state the following lemma, which will be used later in the proof.

\begin{lem}  [Petersen, Pedersen\cite{Petersen08}] \label{lem:VSFG} Assume $\mathbf{x} \backsim \mathcal{N}(\mathbf{m}, \mathbf{\Sigma})$ and $\mathbf{b}, \mathbf{A}$ a vector and a matrix of appropriate dimensions, then
\begin{equation*}
\mathbf{E} \big \lbrace (\mathbf{x} - \mathbf{b})^{T} \mathbf{A} (\mathbf{x} - \mathbf{b}) \big \rbrace = (\mathbf{m} - \mathbf{b})^{T} \mathbf{A} (\mathbf{m} - \mathbf{b}) + \tr{\big( \mathbf{A} \mathbf{\Sigma}\big)}.
\end{equation*}
\end{lem}

\noindent
Next, starting from the DP algorithm given in~(\ref{eq:DP_obs_history_00}), we separately  determine each of the two terms inside the minimization. From the definition of the conditional filtering error covariance matrix in (\ref{eq:ccmfe}) and the filter equation in (\ref{eq:filter_eqn}), we have that
\begin{equation}
\tr{\big (\mathbf{\Sigma}_{k|k}(\mathbf{y}_{k},\mathbf{u}_{k-1})\big )} = 1 - \norm{\mathbf{p}_{k|k-1} + \mathbf{G}_{k} ( \mathbf{y} - \mathbf{y}_{k|k-1} )}^{2}.
\end{equation}

\noindent
Thus, the first term, which corresponds to the immediate cost of selecting control input $\mathbf{u}_{k-1}$, can be computed as follows
\begin{IEEEeqnarray}{rCl}\label{eq:immediate_cost}
\underset{\mathbf{y}_{k}}{\mathbb{E}} \big \lbrace \tr{\big (\mathbf{\Sigma}_{k|k}(\mathbf{y}_{k},\mathbf{u}_{k-1})\big )} \big | \mathcal{F}_{k-1}, \mathbf{u}_{k-1} \big \rbrace 
& = & 
 \int p(\mathbf{y}|\mathcal{F}_{k-1}, \mathbf{u}_{k-1}) \tr{\big (\mathbf{\Sigma}_{k|k}(\mathbf{y},\mathbf{u}_{k-1})\big )} d \mathbf{y} \nonumber \\
& = & 
\sum_{i=1}^{n} p_{k|k-1}^{i} \int f(\mathbf{y} | \mathbf{e}_{i}, \mathbf{u}_{k-1}) \tr{\big (\mathbf{\Sigma}_{k|k}(\mathbf{y},\mathbf{u}_{k-1})\big )} d \mathbf{y} \nonumber \\
& = & 
\sum_{i=1}^{n} p_{k|k-1}^{i} \bigg [ 1 - \mathbb{E} \big \lbrace \norm{\mathbf{p}_{k|k-1} + \mathbf{G}_{k} ( \mathbf{y} - \mathbf{y}_{k|k-1} )}^{2} \big | \mathbf{x}_{k} = \mathbf{e}_{i}, \mathbf{u}_{k-1} \big \rbrace \bigg ]. \IEEEeqnarraynumspace
\end{IEEEeqnarray}

\noindent
To determine the term $\mathbb{E} \big \lbrace \norm{\mathbf{p}_{k|k-1} + \mathbf{G}_{k} [ \mathbf{y} - \mathbf{y}_{k|k-1} ]}^{2} \big | \mathbf{x}_{k} = \mathbf{e}_{i}, \mathbf{u}_{k-1} \big \rbrace$, we work as follows
\begin{IEEEeqnarray}{rCl}\label{eq:intermediate_term_cc}
\mathbb{E} \big \lbrace \norm{\mathbf{p}_{k|k-1} + \mathbf{G}_{k} (\mathbf{y} - \mathbf{y}_{k|k-1})}^{2} \big | \mathbf{x}_{k} = \mathbf{e}_{i}, \mathbf{u}_{k-1} \big \rbrace & = & \norm{\mathbf{p}_{k|k-1}}^{2} + 2 \mathbf{p}_{k|k-1}^{T} \mathbf{E} \big \lbrace \mathbf{G}_{k} (\mathbf{y} -\mathbf{y}_{k|k-1}) | \mathbf{x}_{k} = \mathbf{e}_{i}, \mathbf{u}_{k-1} \big \rbrace \nonumber \\&&+ \mathbb{E} \big \lbrace (\mathbf{y} -\mathbf{y}_{k|k-1})^{T} \mathbf{G}_{k}^{T} \mathbf{G}_{k} (\mathbf{y} -\mathbf{y}_{k|k-1}) | \mathbf{x}_{k} = \mathbf{e}_{i}, \mathbf{u}_{k-1} \big \rbrace \IEEEeqnarraynumspace.
\end{IEEEeqnarray}

\noindent
Note that $\mathbf{G}_{k}$ and $\mathbf{y}_{k|k-1}$ depend by definition on the control input $\mathbf{u}_{k-1}$ and this implies that
\begin{equation}\label{eq:intermediate_term_cc_01}
\mathbf{E} \big \lbrace \mathbf{G}_{k} (\mathbf{y} -\mathbf{y}_{k|k-1}) | \mathbf{x}_{k} = \mathbf{e}_{i}, \mathbf{u}_{k-1} \big \rbrace = \mathbf{G}_{k}(\mathbf{m}_{i}^{\mathbf{u}_{k-1}} - \mathbf{y}_{k|k-1}),
\end{equation}

\noindent
where we have exploited the signal model in (\ref{eq:observation_model}). To determine the $\mathbb{E} \big \lbrace (\mathbf{y} -\mathbf{y}_{k|k-1})^{T} \mathbf{G}_{k}^{T} \mathbf{G}_{k} (\mathbf{y} -\mathbf{y}_{k|k-1}) | \mathbf{x}_{k} = \mathbf{e}_{i}, \mathbf{u}_{k-1} \big \rbrace$, we exploit Lemma \ref{lem:VSFG} and get
\begin{IEEEeqnarray}{rCl}\label{eq:intermediate_term_cc_02}
\mathbb{E} \big \lbrace (\mathbf{y} -\mathbf{y}_{k|k-1})^{T} \mathbf{G}_{k}^{T} \mathbf{G}_{k} (\mathbf{y} -\mathbf{y}_{k|k-1}) | \mathbf{x}_{k} = \mathbf{e}_{i}, \mathbf{u}_{k-1} \big \rbrace & = & (\mathbf{m}_{i}^{\mathbf{u}_{k-1}} - \mathbf{y}_{k|k-1})^{T} \mathbf{G}_{k}^{T} \mathbf{G}_{k} (\mathbf{m}_{i}^{\mathbf{u}_{k-1}} - \mathbf{y}_{k|k-1}) \nonumber \\ &&+ \tr{\big( \mathbf{G}_{k}^{T} \mathbf{G}_{k} \mathbf{\Sigma}_{i}^{\mathbf{u}_{k-1}}\big)}
\end{IEEEeqnarray}

\noindent
Substituting (\ref{eq:intermediate_term_cc_01}) and (\ref{eq:intermediate_term_cc_02}) back to (\ref{eq:intermediate_term_cc}) and combining terms, we get
\begin{equation}
\mathbb{E} \big \lbrace \norm{\mathbf{p}_{k|k-1} + \mathbf{G}_{k} (\mathbf{y} - \mathbf{y}_{k|k-1})}^{2} \big | \mathbf{x}_{k} = \mathbf{e}_{i}, \mathbf{u}_{k-1} \big \rbrace = \norm{\mathbf{p}_{k|k-1} + \mathbf{G}_{k} (\mathbf{m}_{i}^{\mathbf{u}_{k-1}} - \mathbf{y}_{k|k-1})}^{2} + \tr{\big( \mathbf{G}_{k}^{T} \mathbf{G}_{k} \mathbf{\Sigma}_{i}^{\mathbf{u}_{k-1}}\big)}
\end{equation}

\noindent
and the immediate cost of selecting control input $\mathbf{u}_{k-1}$ becomes
\begin{equation}\label{eq:final_im_cost}
\sum_{i=1}^{n} p_{k|k-1}^{i} \big [ 1 -  \tr{\big( \mathbf{G}_{k}^{T} \mathbf{G}_{k} \mathbf{\Sigma}_{i}^{\mathbf{u}_{k-1}}\big)} - \norm{\mathbf{p}_{k|k-1} + \mathbf{G}_{k} (\mathbf{m}_{i}^{\mathbf{u}_{k-1}} - \mathbf{y}_{k|k-1})}^{2} \big ].
\end{equation}

\noindent
The second term in (\ref{eq:DP_obs_history_00}) represents the expected future cost of selecting control input $\mathbf{u}_{k-1}$ and can be determined as follows
\begin{IEEEeqnarray}{rCl}\label{eq:exp_cost}
\underset{\mathbf{y}_{k}}{\mathbb{E}} \big \lbrace J_{k+1}(\mathcal{F}_{k-1},\mathbf{y}_{k},\mathbf{u}_{k-1}) \big | \mathcal{F}_{k-1}, \mathbf{u}_{k-1} \big \rbrace & = & \underset{\mathbf{y}_{k}}{\mathbb{E}} \big \lbrace \overline{J}_{k+1}(\Phi_{k}(\mathbf{p}_{k|k-1},\mathbf{y}_{k},\mathbf{u}_{k-1})) \big | \mathbf{p}_{k|k-1}, \mathbf{u}_{k-1} \big \rbrace \nonumber \\ & = & \int p(\mathbf{y} | \mathbf{p}_{k|k-1}, \mathbf{u}_{k-1})  \overline{J}_{k+1}(\Phi_{k}(\mathbf{p}_{k|k-1},\mathbf{y},\mathbf{u}_{k-1})) d \mathbf{y},
\end{IEEEeqnarray}

\noindent
where we have used the facts that $\mathbf{p}_{k|k-1}$ is a sufficient statistic for $\mathcal{F}_{k-1}$ and $\mathbf{u}_{k-1} = \eta_{k-1}(\mathcal{F}_{k-1})$, and we have denoted by $\Phi_{k}$ the update rule governing the evolution of $\mathbf{p}_{k+1|k}$. At this point, we only need to determine the term $p(\mathbf{y} | \mathbf{p}_{k|k-1}, \mathbf{u}_{k-1})$ and this can be done as follows
\begin{IEEEeqnarray}{rCl}
p(\mathbf{y} | \mathbf{p}_{k|k-1}, \mathbf{u}_{k-1}) = \sum_{i=1}^{n} P(\mathbf{x}_{k}=\mathbf{e}_{i} | \mathbf{p}_{k|k-1}, \mathbf{u}_{k-1}) p(\mathbf{y} | \mathbf{x}_{k}=\mathbf{e}_{i}, \mathbf{u}_{k-1}) = \sum_{i=1}^{n} p_{k|k-1}^{i} f(\mathbf{y} | \mathbf{e}_{i}, \mathbf{u}_{k-1}).
\end{IEEEeqnarray}

\noindent
Substituting back to (\ref{eq:exp_cost}), we get
\begin{IEEEeqnarray}{rCl} \label{eq:final_exp_cost}
\underset{\mathbf{y}_{k}}{\mathbb{E}} \big \lbrace \overline{J}_{k+1}(\Phi_{k}(\mathbf{p}_{k|k-1},\mathbf{y}_{k},\mathbf{u}_{k-1})) \big | \mathbf{p}_{k|k-1}, \mathbf{u}_{k-1} \big \rbrace & = & \int \sum_{i=1}^{n} p_{k|k-1}^{i} f(\mathbf{y} | \mathbf{e}_{i}, \mathbf{u}_{k-1}) \overline{J}_{k+1}(\Phi_{k}(\mathbf{p}_{k|k-1},\mathbf{y},\mathbf{u}_{k-1})) d \mathbf{y} \nonumber \\ & = & \int \mathbf{1}_{n}^{T} \mathbf{r}(\mathbf{y},\mathbf{u}_{k-1}) \mathbf{p}_{k|k-1} \overline{J}_{k+1}\bigg ( \frac{\mathbf{P} \mathbf{r}(\mathbf{y}_{k},\mathbf{u}_{k-1}) \mathbf{p}_{k|k-1}}{\mathbf{1}_{n}^{T} \mathbf{r}(\mathbf{y}_{k},\mathbf{u}_{k-1}) \mathbf{p}_{k|k-1}} \bigg ) d \mathbf{y}, \IEEEeqnarraynumspace
\end{IEEEeqnarray}

\noindent
where we have used the update rule for $\mathbf{p}_{k|k-1}$ given in (\ref{eq:update_rule_vf}). Substituting (\ref{eq:final_im_cost}) and (\ref{eq:final_exp_cost}) to (\ref{eq:DP_obs_history_00}), we get the final form of the DP algorithm given in (\ref{eq:DP_ss_bs_00}). The cost-to-go function for time step $L$ simply consists of the immediate cost of selecting a particular control and has the form given in (\ref{eq:final_im_cost}).

\subsection{Proof of Theorem \ref{thm:Smoother_estimate}}\label{app:pfThm_Smoother_estimate}

The smoothed estimator of $\mathbf{x}_{k}$ can be derived by summing (\ref{eq:smooth_innov_eqn}) from  $s = k$ to $s = R$ and substituting for $\gamma_{s}$ and $\mathbf{C}_{s}$ from (\ref{eq:smooth_innov_eqn}) and (\ref{eq:gain_eqn_smooth_approx}), respectively, as follows
\begin{align}
\sum_{s=k}^{R}\gamma_{s} &= \sum_{s = k}^{R} (\mathbf{p}_{k|s} - \mathbf{p}_{k|s-1}) = \mathbf{p}_{k|R} - \mathbf{p}_{k|k-1} \Rightarrow\\
\mathbf{p}_{k|R} &= \mathbf{p}_{k|k} + \sum_{s=k+1}^{R}\gamma_{s} = \mathbf{p}_{k|k} + \sum_{s = k+1}^{R}\mathbb{E} \lbrace \gamma_{s} \zeta_{s}^{T} | \mathcal{F}_{s-1} \rbrace \big [ \mathbb{E} \lbrace \zeta_{s} \zeta_{s}^{T} | \mathcal{F}_{s-1} \rbrace \big ]^{-1} \zeta_{s}.\label{eq:smooth_formula_general}
\end{align}
 
 \noindent
 At this point, we only need to determine the terms $\mathbb{E} \lbrace \gamma_{s} \zeta_{s}^{T} | \mathcal{F}_{s-1} \rbrace$ and $\mathbb{E} \lbrace \zeta_{s} \zeta_{s}^{T} | \mathcal{F}_{s-1} \rbrace$ and this can be done similarly to the Kalman--like filter gain. The second term can be derived following the same principles as in the derivation of $\mathbb{E} \lbrace \lambda_{k} \lambda_{k}^{T} | \mathcal{F}_{k-1} \rbrace$ and thus, we have
 \begin{equation}\label{eq:smooth_gain_num}
 \mathbb{E} \lbrace \zeta_{s} \zeta_{s}^{T} | \mathcal{F}_{s-1} \rbrace = \widetilde{\mathbf{Q}}_{s} + \mathcal{M}(\mathbf{u}_{s-1}) \mathbf{\Sigma}_{s|s-1} \mathcal{M}^{T}(\mathbf{u}_{s-1}),
 \end{equation}
 
 \noindent
where $ \widetilde{\mathbf{Q}}_{s} = \sum_{i=1}^{n} p_{s|s-1}^{i} \mathbf{Q}_{i}^{\mathbf{u}_{s-1}}$, while for the first term, we work as follows
\begin{IEEEeqnarray}{rCl}
\mathbb{E} \lbrace \gamma_{s} \zeta_{s}^{T} | \mathcal{F}_{s-1} \rbrace & = & \mathbb{E} \lbrace \mathbf{p}_{k|s}\zeta_{s}^{T} | \mathcal{F}_{s-1} \rbrace -\mathbb{E} \lbrace \mathbf{p}_{k|s-1}\zeta_{s}^{T} | \mathcal{F}_{s-1} \rbrace \\ & = &  \mathbb{E} \lbrace \mathbb{E}\lbrace \mathbf{x}_{k} | \mathcal{F}_{s} \rbrace \zeta_{s}^{T} | \mathcal{F}_{s-1} \rbrace - \mathbf{p}_{k|s-1} \mathbb{E} \lbrace \zeta_{s}^{T} | \mathcal{F}_{s-1} \rbrace \\
& \overset{(a)}= & \mathbb{E} \lbrace \mathbb{E}\lbrace \mathbf{x}_{k} \zeta_{s}^{T} | \mathcal{F}_{s} \rbrace | \mathcal{F}_{s-1} \rbrace = \mathbb{E}\lbrace \mathbf{x}_{k} (\mathbf{y}_{s} - \mathbf{y}_{s|s-1})^{T} | \mathcal{F}_{s-1} \rbrace \\
& \overset{(b)}= & \mathbb{E} \lbrace \mathbf{x}_{k} \mathbf{x}_{s}^{T} | \mathcal{F}_{s-1} \rbrace \mathcal{M}^{T}(\mathbf{u}_{s-1}) + \mathbb{E} \lbrace \mathbf{x}_{k} \mathbf{v}_{s}^{T} | \mathcal{F}_{s-1} \rbrace - \mathbf{p}_{k|s-1}\mathbf{y}_{s|s-1}^{T}, \label{eq:smooth_gain_denom}
\end{IEEEeqnarray}

\noindent
where we have exploited that $(a)$ $\zeta_{s}$ is a $\lbrace \mathcal{F} \rbrace$--MD sequence and $(b)$ $\mathbf{u}_{s-1} = \eta_{s-1}(\mathcal{F}_{s-1})$. To derive a closed-form expression for the term $\mathbf{\Theta}_{k,s} = \mathbb{E} \lbrace \mathbf{x}_{k} \mathbf{x}_{s}^{T} | \mathcal{F}_{s-1} \rbrace$, we first observe that
\begin{IEEEeqnarray}{rCl}
\mathbf{\Theta}_{k,s} & = & \mathbb{E} \lbrace \mathbf{x}_{k} \mathbf{x}_{s}^{T} | \mathcal{F}_{s-1} \rbrace = \mathbb{E} \lbrace \mathbf{x}_{k} \mathbf{x}_{s-1}^{T} | \mathcal{F}_{s-1} \rbrace \mathbf{P}^{T} + \mathbb{E} \lbrace \mathbf{x}_{k} \mathbf{w}_{s}^{T} | \mathcal{F}_{s-1} \rbrace \\ & = & \mathbb{E} \lbrace \mathbf{x}_{k} \mathbf{x}_{s-1}^{T} | \mathcal{F}_{s-1} \rbrace \mathbf{P}^{T} + \mathbb{E} \lbrace \mathbb{E} \lbrace \mathbf{x}_{k} \mathbf{w}_{s}^{T} | \mathcal{B}_{s-1} \rbrace | \mathcal{F}_{s-1} \rbrace \\ & = & \mathbb{E} \lbrace \mathbf{x}_{k} \mathbf{x}_{s-1}^{T} | \mathcal{F}_{s-1} \rbrace \mathbf{P}^{T} + \mathbb{E} \lbrace \mathbf{x}_{k} \mathbb{E} \lbrace \mathbf{w}_{s}^{T} | \mathcal{B}_{s-1} \rbrace | \mathcal{F}_{s-1} \rbrace \\ & = & \mathbb{E} \lbrace \mathbf{x}_{k} \mathbf{x}_{s-1}^{T} | \mathcal{F}_{s-1} \rbrace \mathbf{P}^{T}.
\end{IEEEeqnarray}

\noindent
Then, for determining the term $\mathbb{E} \lbrace \mathbf{x}_{k} \mathbf{x}_{s-1}^{T} | \mathcal{F}_{s-1} \rbrace$, we work as follows
\begin{IEEEeqnarray}{rCl}
\mathbb{E} \lbrace \mathbf{x}_{k} \mathbf{x}_{s-1}^{T} | \mathcal{F}_{s-1} \rbrace & = & \sum_{i=1}^{n} \sum_{j=1}^{n} \mathbf{e}_{i} \mathbf{e}_{j}^{T} P(\mathbf{x}_{k} = \mathbf{e}_{i}, \mathbf{x}_{s-1} = \mathbf{e}_{j} | \mathcal{F}_{s-1}) \\
& = & 
\begin{bmatrix}
P(\mathbf{x}_{k} = \mathbf{e}_{1}, \mathbf{x}_{s-1} = \mathbf{e}_{1} | \mathcal{F}_{s-1}) & \dotsc & P(\mathbf{x}_{k} = \mathbf{e}_{1}, \mathbf{x}_{s-1} = \mathbf{e}_{n} | \mathcal{F}_{s-1}) \\
P(\mathbf{x}_{k} = \mathbf{e}_{2}, \mathbf{x}_{s-1} = \mathbf{e}_{1} | \mathcal{F}_{s-1}) & \dotsc & P(\mathbf{x}_{k} = \mathbf{e}_{2}, \mathbf{x}_{s-1} = \mathbf{e}_{n} | \mathcal{F}_{s-1})  \\
\vdots & \ddots & \vdots \\
P(\mathbf{x}_{k} = \mathbf{e}_{n}, \mathbf{x}_{s-1} = \mathbf{e}_{1} | \mathcal{F}_{s-1}) & \dotsc & P(\mathbf{x}_{k} = \mathbf{e}_{n}, \mathbf{x}_{s-1} = \mathbf{e}_{n} | \mathcal{F}_{s-1}) 
\end{bmatrix} \\
& = & \frac{1}{p(\mathbf{y}_{s-1}|\mathcal{F}_{s-2}, \mathbf{u}_{s-2})} \mathbb{E} \lbrace \mathbf{x}_{k} \mathbf{x}_{s-2}^{T} | \mathcal{F}_{s-2} \rbrace \mathbf{P}^{T}
\mathbf{r}(\mathbf{y}_{s-1},\mathbf{u}_{s-2}) \\
& = & \frac{1}{p(\mathbf{y}_{s-1}|\mathcal{F}_{s-2}, \mathbf{u}_{s-2})} \mathbf{\Theta}_{k,s-1}
\mathbf{r}(\mathbf{y}_{s-1},\mathbf{u}_{s-2}) = \frac{\mathbf{\Theta}_{k,s-1}
\mathbf{r}(\mathbf{y}_{s-1},\mathbf{u}_{s-2})}{\mathbf{1}_{n}^{T} [\mathbf{\Theta}_{k,s-1}
\mathbf{r}(\mathbf{y}_{s-1},\mathbf{u}_{s-2})] \mathbf{1}_{n}},
\end{IEEEeqnarray}

\noindent
where $\mathbf{r}(\mathbf{y}_{s-1},\mathbf{u}_{s-2}) = \diag(f(\mathbf{y}_{s-1}|\mathbf{x}_{s-1} = \mathbf{e}_{1}, \mathbf{u}_{s-2}), \dotsc, f(\mathbf{y}_{s-1}|\mathbf{x}_{s-1} = \mathbf{e}_{n}, \mathbf{u}_{s-2}))$ is the $n \times n$ diagonal matrix of measurement vector probability density functions and $\mathbf{1}_{n}$ is a column vector of $n$ ones. Note that the above recursive formula is initialized by $\mathbb{E} \lbrace \mathbf{x}_{0} \mathbf{x}_{0}^{T} | \mathcal{F}_{0} \rbrace = \diag(\mathbf{p}_{0|0})$. For the term $\mathbb{E} \lbrace \mathbf{x}_{k} \mathbf{v}_{s}^{T} | \mathcal{F}_{s-1} \rbrace$, we have
\begin{equation}\label{eq:smooth_gain_denom_term_2}
\mathbb{E} \lbrace \mathbf{x}_{k} \mathbf{v}_{s}^{T} | \mathcal{F}_{s-1} \rbrace = \mathbb{E} \lbrace \mathbb{E} \lbrace \mathbf{x}_{k} \mathbf{v}_{s}^{T} | \mathcal{B}_{s}^{-} \rbrace | \mathcal{F}_{s-1} \rbrace = \mathbb{E} \lbrace \mathbf{x}_{k} \mathbb{E} \lbrace  \mathbf{v}_{s}^{T} | \mathcal{B}_{s}^{-} \rbrace | \mathcal{F}_{s-1} = 0,
\end{equation}

\noindent
where we have exploited the MD property of $\mathbf{v}_{s}$ along with the fact that $\mathbf{x}_{k} \in \mathcal{B}_{s}^{-}, \forall s > k$. The above results for $\mathbf{\Theta}_{k,s}$ and $\mathbb{E} \lbrace \mathbf{x}_{k} \mathbf{v}_{s}^{T} | \mathcal{F}_{s-1} \rbrace$ allow us to rewrite (\ref{eq:smooth_gain_denom}) as
\begin{equation}\label{eq:smooth_gain_denom_final}
\mathbb{E} \lbrace \gamma_{s} \zeta_{s}^{T} | \mathcal{F}_{s-1} \rbrace = \mathbf{\Theta}_{k,s}\mathcal{M}^{T}(\mathbf{u}_{s-1}) - \mathbf{p}_{k|s-1} \mathbf{y}_{s|s-1}^{T} = (\mathbf{\Theta}_{k,s} - \mathbf{p}_{k|s-1} \mathbf{p}_{s|s-1}^{T})\mathcal{M}^{T}(\mathbf{u}_{s-1})
\end{equation}

\noindent
Finally, substituting (\ref{eq:smooth_gain_num}) and (\ref{eq:smooth_gain_denom_final}) back to (\ref{eq:gain_eqn_smooth_approx}) completes the proof.

\subsection{Proof of Proposition \ref{prp:fi_smoother}}\label{app:pf_fi_smoother}

The fixed--point smoother is defined as
\begin{equation}\label{eq:fi_smoother_deriv_1}
\mathbf{p}_{k|L} = \mathbf{p}_{k|k} + \sum_{s = k+1}^{L} \mathbf{C}_{s} (\mathbf{y}_{s} - \mathbf{y}_{s|s-1})
\end{equation}

\noindent
and setting $k = k - 1$, we get
\begin{equation}\label{eq:fi_smoother_deriv_2}
\mathbf{p}_{k-1|L} = \mathbf{p}_{k-1|k-1} + \sum_{s = k}^{L} \mathbf{C}_{s} (\mathbf{y}_{s} - \mathbf{y}_{s|s-1}).
\end{equation}

\noindent
Multiplying (\ref{eq:fi_smoother_deriv_2}) by $\mathbf{P}$, subtracting it from (\ref{eq:fi_smoother_deriv_1}) and rearranging terms, gives us the following
\begin{equation}\label{eq:fi_smoother_deriv_3}
\mathbf{p}_{k|L} = \mathbf{P} \mathbf{p}_{k-1|L} + \mathbf{p}_{k|k} - \mathbf{p}_{k|k-1} - \mathbf{P} \mathbf{G}_{k} (\mathbf{y}_{k} - \mathbf{y}_{k|k-1}) + (\mathbf{I}_{n} - \mathbf{P}) \sum_{s = k+1}^{L} \mathbf{C}_{s}(\mathbf{y}_{s} - \mathbf{y}_{s|s-1}).
\end{equation}

\noindent
At this point, we observe from the filter definition in (\ref{eq:filter_eqn}) that
\begin{equation}\label{eq:fi_smoother_deriv_4}
\mathbf{p}_{k|k} - \mathbf{p}_{k|k-1} - \mathbf{P} \mathbf{G}_{k} (\mathbf{y}_{k} - \mathbf{y}_{k|k-1}) = (\mathbf{I}_{n} - \mathbf{P}) \mathbf{C}_{k} (\mathbf{y}_{k} - \mathbf{y}_{k|k-1}),
\end{equation}

\noindent
where we have exploited that $\mathbf{G}_{k}$ coincides with $\mathbf{C}_{k}$. Substituting (\ref{eq:fi_smoother_deriv_4}) back to (\ref{eq:fi_smoother_deriv_3}) gives us the final form of the fixed--interval smoother given in (\ref{eq:fi_smoother}).

\subsection{Proof of Proposition \ref{prp:fl_smoother}}\label{app:pf_fl_smoother}

Setting $L = k + \Delta$ in the formula for the fixed--point smoother, we get
\begin{equation}\label{eq:fl_smoother_deriv_1}
\mathbf{p}_{k|k+\Delta} = \mathbf{p}_{k|k} + \sum_{s = k+1}^{k+\Delta} \mathbf{C}_{s} (\mathbf{y}_{s} - \mathbf{y}_{s|s-1}),
\end{equation}

\noindent
and for $k = k-1$, (\ref{eq:fl_smoother_deriv_1}) becomes
\begin{equation}\label{eq:fl_smoother_deriv_2}
\mathbf{p}_{k-1|k+\Delta-1} = \mathbf{p}_{k-1|k-1} + \sum_{s = k}^{k+\Delta-1} \mathbf{C}_{s} (\mathbf{y}_{s} - \mathbf{y}_{s|s-1}).
\end{equation}

\noindent
Multiplying (\ref{eq:fl_smoother_deriv_2}) by $\mathbf{P}$, subtracting it from (\ref{eq:fl_smoother_deriv_1}) and rearranging terms, gives us the following
\begin{equation}
\begin{split}
\mathbf{p}_{k|k+\Delta} = & ~\mathbf{P} \mathbf{p}_{k-1|k+\Delta-1} + [\mathbf{C}_{k+\Delta} (\mathbf{y}_{k+\Delta} - \mathbf{y}_{k+\Delta | k + \Delta - 1}) - \mathbf{P} \mathbf{C}_{k} ( \mathbf{y}_{k} - \mathbf{y}_{k|k-1}) - \mathbf{P}\mathbf{p}_{k-1|k-1} ] \\&+ (\mathbf{I}_{n} - \mathbf{P}) \sum_{s = k+1}^{k+\Delta-1} \mathbf{C}_{s} (\mathbf{y}_{s} - \mathbf{y}_{s|s-1}).
\end{split}
\end{equation}

\noindent
We observe that
\begin{equation}
\mathbf{P} \mathbf{C}_{k} ( \mathbf{y}_{k} + \mathbf{y}_{k|k-1}) - \mathbf{P}\mathbf{p}_{k-1|k-1} = \mathbf{p}_{k+1|k} + \mathbf{p}_{k|k-1} - \mathbf{P} \mathbf{p}_{k|k-1},
\end{equation}

\noindent
and after setting $\Gamma(k,\Delta) = \mathbf{C}_{k+\Delta} (\mathbf{y}_{k+\Delta} - \mathbf{y}_{k+\Delta | k + \Delta - 1}) - \mathbf{p}_{k+1|k} - \mathbf{p}_{k|k-1} + \mathbf{P} \mathbf{p}_{k|k-1}$, we obtain the final form of the fixed--lag smoother given in (\ref{eq:fl_smoother}).

\end{appendix}

\bibliographystyle{IEEEtran}
\bibliography{IEEEabrv,references}

\begin{thebibliography}{10}
\providecommand{\url}[1]{#1}
\csname url@samestyle\endcsname
\providecommand{\newblock}{\relax}
\providecommand{\bibinfo}[2]{#2}
\providecommand{\BIBentrySTDinterwordspacing}{\spaceskip=0pt\relax}
\providecommand{\BIBentryALTinterwordstretchfactor}{4}
\providecommand{\BIBentryALTinterwordspacing}{\spaceskip=\fontdimen2\font plus
\BIBentryALTinterwordstretchfactor\fontdimen3\font minus
  \fontdimen4\font\relax}
\providecommand{\BIBforeignlanguage}[2]{{%
\expandafter\ifx\csname l@#1\endcsname\relax
\typeout{** WARNING: IEEEtran.bst: No hyphenation pattern has been}%
\typeout{** loaded for the language `#1'. Using the pattern for}%
\typeout{** the default language instead.}%
\else
\language=\csname l@#1\endcsname
\fi
#2}}
\providecommand{\BIBdecl}{\relax}
\BIBdecl

\bibitem{WilliamsTSP07}
J.~L. Williams, J.~W. Fisher, and A.~S. Willsky, ``Approximate {D}ynamic
  {P}rogramming for {C}ommunication-{C}onstrained {S}ensor {N}etwork
  {M}anagement,'' \emph{IEEE Transactions on Signal Processing,}, vol.~55,
  no.~8, pp. 4300--4311, August 2007.

\bibitem{AtiaTSP11}
G.~K. Atia, V.~V. Veeravalli, and J.~A. Fuemmeler, ``Sensor {S}cheduling for
  {E}nergy-{E}fficient {T}arget {T}racking in {S}ensor {N}etworks,'' \emph{IEEE
  Transactions on Signal Processing}, vol.~59, no.~10, pp. 4923--4937, October
  2011.

\bibitem{HeroSJ11}
A.~O. Hero and D.~Cochran, ``Sensor {M}anagement: {P}ast, {P}resent, and
  {F}uture,'' \emph{IEEE Sensors Journal}, vol.~11, no.~12, pp. 3064--3075,
  2011.

\bibitem{NaghshvararXiv13}
M.~Naghshvar, T.~Javidi, and M.~Wigger, ``Extrinsic {J}ensen--{S}hannon
  {D}ivergence: {A}pplications to {V}ariable-{L}ength {C}oding,'' June 2013,
  arXiv:1307.0067.

\bibitem{UnnikrishnanTSP10}
J.~Unnikrishnan and V.~V. Veeravalli, ``Algorithms for {D}ynamic {S}pectrum
  {A}ccess with {L}earning for {C}ognitive {R}adio,'' \emph{IEEE Transactions
  on Signal Processing}, vol.~58, no.~2, pp. 750--760, February 2010.

\bibitem{RangarajanJSTSP07}
R.~Rangarajan, R.~Raich, and A.~O. Hero, ``Optimal {S}equential {E}nergy
  {A}llocation for {I}nverse {P}roblems,'' \emph{IEEE Journal of Selected
  Topics in Signal Processing}, vol.~1, no.~1, pp. 67--78, June 2007.

\bibitem{NaghshvarISIT13}
M.~Naghshvar and T.~Javidi, ``Two-dimensional visual search,'' in
  \emph{International Symposium Information Theory (ISIT)}, July 2013.

\bibitem{HauptSSP12}
J.~Haupt, R.~Baraniuk, R.~Castro, and R.~Nowak, ``Sequentially designed
  compressed sensing,'' in \emph{IEEE Statistical Signal Processing Workshop
  (SSP)}, August 2012, pp. 401--404.

\bibitem{WeiJSTSP13}
D.~Wei and A.~O. Hero, ``Multistage {A}daptive {E}stimation of {S}parse
  {S}ignals,'' \emph{IEEE Journal of Selected Topics in Signal Processing},
  vol.~7, no.~5, pp. 783--796, April 2013.

\bibitem{MalloyarXiv13}
M.~L. Malloy and R.~D. Nowak, ``Near-{O}ptimal {A}daptive {C}ompressed
  {S}ensing,'' June 2013, arXiv:1306.6239.

\bibitem{KrishnamurthyTSP12}
V.~Krishnamurthy, R.~R. Bitmead, M.~Gevers, and E.~Miehling, ``Sequential
  {D}etection with {M}utual {I}nformation {S}topping {C}ost,'' \emph{IEEE
  Transactions on Signal Processing}, vol.~60, no.~2, pp. 700--714, February
  2012.

\bibitem{RangangaranICASSP05}
R.~Rangarajan, R.~Raich, and A.~O. Hero, ``Optimal experimental design for an
  inverse scattering problem,'' in \emph{IEEE International Conference on
  Acoustics, Speech, and Signal Processing (ICASSP)}, vol.~4, March 2005.

\bibitem{LigoICASSP13}
J.~Ligo, G.~Atia, and V.~V. Veeravalli, ``A {C}ontrolled {S}ensing {A}pproach
  to {G}raph {C}lassification,'' in \emph{IEEE International Conference on
  Acoustics, Speech, and Signal Processing (ICASSP)}, May 2013.

\bibitem{ZoisTSP13}
D.-S. Zois, M.~Levorato, and U.~Mitra, ``Energy--{E}fficient, {H}eterogeneous
  {S}ensor {S}election for {P}hysical {A}ctivity {D}etection in {W}ireless
  {B}ody {A}rea {N}etworks,'' \emph{IEEE Tansactions on Signal Processing},
  vol.~61, no.~7, pp. 1581--1594, April 2013.

\bibitem{RiccardiTSAP05}
G.~Riccardi and D.~Hakkani-Tur, ``Active learning: theory and applications to
  automatic speech recognition,'' \emph{IEEE Transactions on Speech and Audio
  Processing}, vol.~13, no.~4, pp. 504--511, 2005.

\bibitem{NowakIT11}
R.~D. Nowak, ``The geometry of generalized binary search,'' \emph{IEEE
  Transactions on Information Theory}, vol.~57, pp. 7893--7906, December 2011.

\bibitem{BertsekasDPOC05}
D.~P. Bertsekas, \emph{Dynamic {P}rogramming and {O}ptimal {C}ontrol}.\hskip
  1em plus 0.5em minus 0.4em\relax Athena Scientific, 2005, vol.~1.

\bibitem{ZoisGlobalSIP13}
D.-S. Zois and U.~Mitra, ``On the {P}roperties of {N}onlinear {POMDPs} for
  {A}ctive {S}tate {T}racking,'' in \emph{1st IEEE Global Conference on Signal
  and Information Processing (GlobalSIP)}, December 2013.

\bibitem{MaggioniTR06}
M.~Maggioni and S.~Mahadevan, ``A multiscale framework for {M}arkov decision
  processes using diffusion wavelets,'' University of Massachusetts, Tech.
  Rep., 2006.

\bibitem{LevoratoEurasip12}
M.~Levorato, U.~Mitra, and A.~Goldsmith, ``Structure-based learning in wireless
  networks via sparse approximation,'' \emph{EURASIP Journal on Wireless
  Communications and Networking}, vol. 278, no.~1, pp. 1--15, August 2012.

\bibitem{ChenTR03}
Z.~Chen, ``Bayesian {F}iltering: {F}rom {K}alman {F}ilters to {P}article
  {F}ilters, and {B}eyond,'' McMaster University, Tech. Rep., 2003.

\bibitem{RauchJAIAA65}
H.~E. Rauch, F.~Tung, and C.~T. Striebel, ``Maximum likelihood estimates of
  linear dynamic systems,'' \emph{J. Amer. Inst. Aeronautics and Astronautics},
  vol.~3, no.~8, pp. 1445–--1450, August 1965.

\bibitem{MeditchJIC67}
J.~S. Meditch, ``On optimal linear smoothing theory,'' \emph{J. Inform.
  Contr.}, vol.~10, pp. 598--615, August 1967.

\bibitem{MooreA73}
J.~B. Moore, ``Discrete-time fixed-lag smoothing algorithms,''
  \emph{Automatica}, vol.~9, no.~2, pp. 163--174, 1973.

\bibitem{Haykin01}
E.~A. Wan and R.~Van Der~Merwe, ``The unscented {K}alman filter,'' in
  \emph{Kalman Filtering and Neural Networks}, S.~Haykin, Ed.\hskip 1em plus
  0.5em minus 0.4em\relax Wiley, 2001.

\bibitem{RibeiroTSP06}
A.~Ribeiro, G.~B. Giannakis, and S.~I. Roumeliotis, ``{SOI-KF}: {D}istributed
  {K}alman {F}iltering {W}ith {L}ow-{C}ost {C}ommunications {U}sing the {S}ign
  of {I}nnovations,'' \emph{IEEE Transactions on Signal Processing}, vol.~54,
  no.~12, pp. 4782--4795, Dec. 2006.

\bibitem{RibeiroCSM10}
A.~Ribeiro, I.~D. Schizas, S.~Roumeliotis, and G.~B. Giannakis, ``Kalman
  {F}iltering in {W}ireless {S}ensor {N}etworks,'' \emph{IEEE Control Systems},
  vol.~30, no.~2, pp. 66--86, 2010.

\bibitem{NerurkarTR12}
E.~D. Nerurkar and S.~I. Roumeliotis, ``Resource-aware {H}ybrid {E}stimation
  {F}ramework for {M}ulti-robot {C}ooperative {L}ocalization,'' University of
  Minnesota, Tech. Rep. 2012-001, 2012.

\bibitem{SegallIT76SPET}
A.~Segall, ``Stochastic {P}rocesses in estimation theory,'' \emph{IEEE
  Transactions on Information Theory}, vol.~22, no.~3, pp. 275--286, May 1976.

\bibitem{Bremaud81}
P.~Bremaud, \emph{Point {P}rocesses and {Q}ueues}.\hskip 1em plus 0.5em minus
  0.4em\relax Springer, 1981.

\bibitem{ElliottSV95}
R.~J. Elliott, L.~Aggoun, and J.~B. Moore, \emph{Hidden {M}arkov {M}odels,
  {E}stimation and {C}ontrol}, ser. Applications of Mathematics.\hskip 1em plus
  0.5em minus 0.4em\relax Springer, 1995, vol.~29.

\bibitem{SegallIT76REDTPP}
A.~Segall, ``Recursive estimation from discrete-time point processes,''
  \emph{IEEE Transactions on Information Theory}, vol.~22, no.~4, pp. 422--431,
  Sep. 1976.

\bibitem{ElliottAMO94}
R.~J. Elliott and H.~Yang, ``How to count and guess well: Discrete adaptive
  filters,'' \emph{Applied Mathematics \& Optimization}, vol.~30, no.~1, pp.
  51--78, July 1994.

\bibitem{BaccarelliSP96}
E.~Baccarelli and R.~Cusani, ``Recursive {K}alman-type optimal estimation and
  detection of hidden {M}arkov chains,'' \emph{Signal Process.}, vol.~51,
  no.~1, pp. 55--64, May 1996.

\bibitem{ElliottA94}
R.~J. Elliott, ``Exact adaptive filters for {M}arkov chains observed in
  {G}aussian noise,'' \emph{Automatica}, vol.~30, no.~9, pp. 1399--1408, Sep.
  1994.

\bibitem{KrishnamurthyTSP93}
V.~Krishnamurthy and J.~B. Moore, ``On-{L}ine {E}stimation of {H}idden {M}arkov
  {M}odel {P}arameters based on the {K}ullback-{L}eibler {I}nformation
  {M}easure,'' \emph{IEEE Transactions on Signal Processing}, vol.~41, no.~8,
  pp. 2557--2573, August 1993.

\bibitem{PhamdoIT94}
N.~Phamdo and N.~Farvardin, ``Optimal {D}etection of {D}iscrete {M}arkov
  {S}ources {O}ver {D}iscrete {M}emoryless {C}hannels--{A}pplications to
  {C}ombined {S}ource-{C}hannel {C}oding,'' \emph{IEEE Transactions on
  Information Theory}, vol.~40, no.~1, pp. 186--193, Jan. 1994.

\bibitem{DeySCL95}
S.~Dey and J.~B. Moore, ``Risk-sensitive filtering and smoothing for {H}idden
  {M}arkov {M}odels,'' \emph{Systems and Control Letters}, vol.~25, no.~5, pp.
  361--366, August 1995.

\bibitem{AckersonTAC70}
G.~A. Ackerson and K.~S. Fu, ``On state estimation in switching environments,''
  \emph{IEEE Transactions on Automatic Control}, vol.~15, no.~1, pp. 10–--17,
  February 1970.

\bibitem{TugnaitAutomatica82}
J.~K. Tugnait, ``Detection and estimation for abruptly changing systems,''
  \emph{Automatica}, vol.~18, no.~5, pp. 607--615, 1982.

\bibitem{CostaDTMJLS05}
O.~L.~V. Costa, M.~D. Fragoso, and R.~P. Marques, \emph{Discrete-Time Markov
  Jump Linear Systems}.\hskip 1em plus 0.5em minus 0.4em\relax Springer
  Probability and its Applications Series, 2005.

\bibitem{BlackmoreTAC08}
L.~Blackmore, S.~Rajamanoharan, and B.~C. Williams, ``Active {E}stimation for
  {J}ump {M}arkov {L}inear {S}ystems,'' \emph{IEEE Transactions on Automatic
  Control}, vol.~53, no.~10, pp. 2223--2236, 2008.

\bibitem{KrishnamurthyTSP02}
V.~Krishnamurthy, ``Algorithms for optimal scheduling and management of hidden
  {M}arkov model sensors,'' \emph{IEEE Transactions on Signal Processing},
  vol.~50, no.~6, pp. 1382--1397, June 2002.

\bibitem{GuptaAutomatica06}
V.~Gupta, T.~H. Chung, B.~Hassibi, and R.~M. Murray, ``On a stochastic sensor
  selection algorithm with applications in sensor scheduling and sensor
  coverage,'' \emph{Automatica}, vol.~42, no.~2, pp. 251--260, February 2006.

\bibitem{KrishnamurthyTSP07}
V.~Krishnamurthy and D.~Djonin, ``Structured {T}hreshold {P}olicies for
  {D}ynamic {S}ensor {S}cheduling -- a {P}artially {O}bserved {M}arkov
  {D}ecision {P}rocess {A}pproach,'' \emph{IEEE Transactions on Signal
  Processing}, vol.~55, no.~10, pp. 4938--4957, October 2007.

\bibitem{NaghshvararXiv12}
M.~Naghshvar and T.~Javidi, ``Active {S}equential {H}ypothesis {T}esting,''
  October 2012, arXiv:1203.4626v3.

\bibitem{MasazadeCISS12}
E.~Masazade, R.~Niu, and P.~K. Varshney, ``An approximate dynamic programming
  based non-myopic sensor selection method for target tracking,'' in \emph{46th
  Annual Conference on Information Sciences and Systems (CISS)}, March 2012,
  pp. 1--6.

\bibitem{WuTAC08}
W.~Wu and A.~Arapostathis, ``Optimal {S}ensor {Q}uerying: {G}eneral {M}arkovian
  and {LQG} {M}odels with {C}ontrolled {O}bservations,'' \emph{IEEE
  Transactions on Automatic Control}, vol.~53, no.~6, pp. 1392--1405, July
  2008.

\bibitem{KrishnamurthyIT13}
V.~Krishnamurthy, ``How to {S}chedule {M}easurements of a {N}oisy {M}arkov
  {C}hain in {D}ecision {M}aking?'' \emph{IEEE Transactions on Information
  Theory}, vol.~59, no.~7, pp. 4440--4461, March 2013.

\bibitem{NitinawaratTAC13}
S.~Nitinawarat, G.~K. Atia, and V.~V. Veeravalli, ``Controlled {S}ensing for
  {M}ultihypothesis {T}esting,'' \emph{IEEE Transactions on Automatic Control},
  vol.~58, no.~10, pp. 2451--2464, October 2013.

\bibitem{FedorovTOE72}
V.~Fedorov, \emph{Theory of {O}ptimal {E}xperiments}.\hskip 1em plus 0.5em
  minus 0.4em\relax New York: Academic, 1972.

\bibitem{SettlesTechRep09}
B.~Settles, ``Active {L}earning {L}iterature {S}urvey,'' University of
  Wisconsin--Madison, Computer Sciences Technical Report 1648, 2009.

\bibitem{GolovinNIPS10}
D.~Golovin, A.~Krause, and D.~Ray, ``Near-{O}ptimal {B}ayesian {A}ctive
  {L}earning with {N}oisy {O}bservations,'' in \emph{2010 Neural Information
  Processing Systems (NIPS)}, 2010, pp. 766--774.

\bibitem{NaghshvarAllerton12}
M.~Naghshvar, T.~Javidi, and C.~Chaudhuri, ``Noisy {B}ayesian {A}ctive
  {L}earning,'' in \emph{Fiftieth Annual Allerton Conference Allerton House},
  October 2012.

\bibitem{MarcusTAC79}
S.~I. Marcus, ``Optimal {N}onlinear {E}stimation for a {C}lass of
  {D}iscrete-{T}ime {S}tochastic {S}ystems,'' \emph{IEEE Transactions on
  Automatic Control}, vol.~24, no.~2, pp. 297--302, 1979.

\bibitem{BremaudTR76a}
P.~M. Bremaud and J.~H. Van~Schuppen, ``Discrete time stochastic systems: Part
  {I}: stochastic calculus and representations,'' Tech. Rep. SSM 7603a, June
  1976.

\bibitem{BremaudTR76b}
------, ``Discrete time processes: Part {II}: estimation theory,'' Tech. Rep.
  SSM 7603b, June 1976.

\bibitem{GalliTC02}
S.~Galli, ``A new family of soft-output adaptive receivers exploiting nonlinear
  {MMSE} estimates for {TDMA}-based wireless links,'' \emph{IEEE Transactions
  on Communications}, vol.~50, no.~12, pp. 1935--1945, Dec. 2002.

\bibitem{SpeyerChung08}
J.~L. Speyer and W.~H. Chung, \emph{Stochastic {P}rocesses, {E}stimation and
  {C}ontrol}.\hskip 1em plus 0.5em minus 0.4em\relax SIAM, 2008.

\bibitem{LovejoyAOR91}
W.~S. Lovejoy, ``A survey of algorithmic methods for partially observed
  {M}arkov decision processes,'' \emph{Annals of Operations Research}, vol.~28,
  pp. 47--66, April 1991.

\bibitem{ThatteTSP11}
G.~Thatte, M.~Li, S.~Lee, B.~A. Emken, M.~Annavaram, S.~Narayanan,
  D.~Spruijt-Metz, and U.~Mitra, ``Optimal {T}ime-{R}esource {A}llocation for
  {E}nergy-{E}fficient {P}hysical {A}ctivity {D}etection,'' \emph{IEEE
  Tansactions on Signal Processing}, vol.~59, no.~4, pp. 1843 --1857, April
  2011.

\bibitem{Petersen08}
K.~B. Petersen and P.~M. S., ``The {M}atrix {C}ookbook,'' November 2008.

\end{thebibliography}


\begin{table}[h!]
\caption{Detection accuracy for different control policies (A: ACC mean -- 1 sample, B: ACC variance -- 1 sample,  $\Gamma$: ECG Period -- 1 sample.)}
\begin{center}
  \begin{tabular}{ | C{1.4in} || c | c | c | c |}
    \hline
    Control policy & A & B & $\Gamma$ & Optimal \\ \hline
    Detection accuracy & $74\%$ & $77\%$ & $40\%$ & $85\%$ \\ \hline
  \end{tabular}
\end{center}
\label{tb:detection_accuracy}
\end{table}

\begin{table}[h!]
\caption{Filtering and smoothing detection accuracy.}
   \label{tab:detection_accuracy}
   \centering
   \begin{tabular}{|c||c|}
     \hline
     Filtering & $85\%$\\
     \hline
     Smoothing, $R = k+1$ & $87\%$\\
     \hline
     Smoothing, $R = k+2$ & $88\%$\\
     \hline
     Smoothing, $R = k+3$ & $88.2\%$\\
     \hline
     Smoothing, $R = k+4$ & $88.4\%$\\
     \hline
   \end{tabular}
\end{table}

\end{document}